\documentclass[a4paper,12pt]{article}

\usepackage{amssymb,cite}%cite is in clash with hyperref!
\newif\ifiTheta
\iThetafalse   % convention without the i
\newcommand{\ft}[2]{{\textstyle\frac{#1}{#2}}}

\def\trace{\mathop{\rm Tr}\nolimits}
\def\rmi{{\rm i}}
\def\rmd{{\rm d}}
\def\rme{{\rm e}}
\newsavebox{\uuunit}
\sbox{\uuunit}
    {\setlength{\unitlength}{0.825em}
     \begin{picture}(0.6,0.7)
        \thinlines
        \put(0,0){\line(1,0){0.5}}
        \put(0.15,0){\line(0,1){0.7}}
        \put(0.35,0){\line(0,1){0.8}}
       \multiput(0.3,0.8)(-0.04,-0.02){12}{\rule{0.5pt}{0.5pt}}
     \end {picture}}
\newcommand {\unity}{\mathord{\!\usebox{\uuunit}}}
\csname @addtoreset\endcsname{equation}{section}
\newcommand{\upomega}{\mbox{\usefont{U}{psy}{m}{n}w}}

\begin{document}

 %%%%%%%%%%%%%%%%%%%%%%%%%%%%%%%%%%%%%%%%%%%%%%%%%%%%%%%%%%%
\begin{titlepage}
\begin{flushright}
KUL-TF-04/29\\
% HU-EP 04/XX\\
SU-ITP-04/034\\
hep-th/0410126
\end{flushright}
\vspace{.5cm}
\begin{center}
\baselineskip=16pt {\bf \LARGE On the fakeness of fake supergravity
}\\
\vfill%\vskip 10mm%27.mm
{\large
Alessio Celi$^1$, Anna Ceresole$^2$, Gianguido Dall'Agata$^3$,\\ \vspace{6pt}
Antoine Van Proeyen$^1$
and Marco Zagermann$^4$
} \\
\vfill%\vskip 7mm%1cm
{\small
$^1$Instituut voor Theoretische Fysica, Katholieke Universiteit Leuven,\\
       Celestijnenlaan 200D B-3001 Leuven, Belgium.\\ \vspace{6pt}
$^2$ INFN, Sezione di Torino and
\\ Dipartimento di Fisica Teorica, Universit{\`a} di Torino \\
Via P. Giuria 1,  I-10125 Torino, Italy.\\ \vspace{6pt}
$^3$ Institut f{\"u}r Physik, Humboldt Universit{\"a}t  \\
Newtonstrasse 15, 12489 Adlershof, Berlin, Germany.   \\
{}From 1st October: Theory Unit, CERN, Geneva 23, CH1211,
Switzerland\\
\vspace{6pt}
$^4$ Department of Physics, Stanford University,\\
Varian Building, Stanford, CA 94305-4060, USA.\\ \vspace{6pt}
 }
\end{center}
\vfill
\begin{center}
{\bf Abstract}
\end{center}
{\small We revisit and complete the study of curved BPS-domain walls in
matter-coupled $5D$, $\mathcal{N}=2$ supergravity and carefully analyse
the relation to gravitational theories known as ``fake supergravities''.
We first show that \emph{curved} BPS-domain walls require the presence of
non-trivial hypermultiplet scalars, whereas walls that are solely
supported by vector multiplet scalars  are necessarily \emph{flat}, due
to the constraints from very special geometry. We then recover fake
supergravity as the effective description of true supergravity where one
restricts the attention to the flowing scalar field  of a given
BPS-domain wall. In general, however, true supergravity  can be simulated
by fake supergravity at most \emph{locally}, based upon two choices: (i)
a suitable adapted coordinate system on the scalar manifold, such that
only one scalar field plays a dynamical r\^{o}le, and (ii) a gauge fixing
of the $SU(2)$ connection on the quaternionic-K{\"a}hler manifold, as this
connection does not fit the simple formalism of fake supergravity.
Employing these gauge and coordinate choices, the BPS-equations for both
vector and hypermultiplet scalars become identical to the fake
supergravity equations, once the line of flow is determined by the full
supergravity equations.

 }\vspace{2mm} \vfill \hrule width 3.cm
{\footnotesize \noindent e-mails: Alessio.Celi@fys.kuleuven.ac.be,
ceresole@to.infn.it, Gianguido.Dall'Agata@cern.ch,\\
Antoine.VanProeyen@fys.kuleuven.ac.be, zagermann@itp.stanford.edu    }
\end{titlepage}
\tableofcontents{}
%\newpage
%%%%%%%%%%%%%%%%

\baselineskip 6 mm

\section{Introduction}
The study of domain wall solutions of $(d+1)$-dimensional (super-)gravity
theories has been an active area of research over the past few   years.
This research is  largely driven by
 applications in the context of the AdS/CFT correspondence
and certain  brane world models.

Most of the domain walls that have been studied in this context are
``Minkowski--sliced'' (or ``flat'' or  ``planar'') domain walls. That is,
having a metric of the form
\begin{equation}
\rmd s^2=\rme^{2U(r)}\eta_{mn}\,\rmd x^m \,\rmd x^n + \rmd r^2
\end{equation}
with $\eta_{mn}=\textrm{diag}(-1,1,\ldots,1)$, they preserve the
isometries of the  $d$-dimensional Poincar{\'e} group. When these domain
walls are supersymmetric and non-singular, one expects them to be stable
solutions of the underlying supergravity theory, based on standard
arguments that involve the existence of  Killing spinors and the
first-order form of the BPS equations.

The stability arguments for flat BPS  domain walls can be   formalized
and extended to theories that are not necessarily supersymmetric
\cite{Townsend:1984iu,Skenderis:1999mm}. In this approach, the classical
stability of a solution is proven by defining a  spinor energy  along the
lines of \cite{Witten:1981mf,JN} by using some \emph{formal}
``transformation laws''
that encode the equations of motion in a first-order form.
 These formal transformation laws have a structure
similar  to the supersymmetry transformation laws in true %real
supergravity theories. For this to be possible, one needs to find  a
scalar function $W(\phi)$ of the scalar fields which  is related to the
scalar  potential $V(\phi)$ in the same way the superpotential is related
to the scalar potential in true supergravity. Such a function $W(\phi)$
is often called an \emph{adapted} superpotential. For the stability
argument to work, however, this only needs to be a \emph{formal} analogy:
the function $W(\phi)$ (provided it exists), need not be a genuine
superpotential of a true supergravity theory. In order to emphasize that
it is in general only a formal analogy to genuine supergravity, this
formalism has been named ``fake supergravity'' \cite{Freedman:2003ax}.

In \cite{Freedman:2003ax}, an attempt  was also made  to generalize these
fake supergravity arguments for classical stability to ``curved'', or
more precisely, to ``$AdS_d$--sliced'' domain walls, i.e., to domain
walls of the form
\begin{equation}
\rmd s^2=\rme^{2U(r)}g_{mn}(x)\,\rmd x^m \,\rmd x^n + \rmd r^2
\label{curvedmetric}
\end{equation}
with $g_{mn}(x)$ being a metric of  $AdS_{d}$ with curvature scale
$L_{d}$. In order to do so, the authors of \cite{Freedman:2003ax}
promoted  the scalar function $W(\phi)$ to an $su(2)$-valued
   $(2 \times 2)$-matrix  $\mathbf{W}(\phi)=W_{i}{}^j(\phi)$
   $(i,j=1,2)$, such that the usual formula defining the scalar
potential $V(\phi)$ is given by
\begin{equation}
V(\phi)=\frac{2(d-1)^2}{\kappa^2}\left(\frac12\trace\right) \left[
\frac{1}{\kappa^2}(\partial _\phi \mathbf{W})^2
-\frac{d}{d-1}\mathbf{W}^{2} \right]. \label{Freedman0a}
\end{equation}
The matrix $\mathbf{W}$ also substitutes the scalar superpotential $W$ in
the corresponding ``fake'' Killing spinor equations for an
$SU(2)$-doublet spinor $\epsilon$:
\begin{eqnarray} \left[ \nabla_{\mu}
+\gamma_{\mu} \mathbf{W} \right] \epsilon &=& 0, \nonumber\\
[2mm]
 \left[ \gamma^\mu \nabla_\mu \phi - \frac{2(d-1)}{\kappa^2}\partial _\phi
 \mathbf{W}\right] \epsilon &=& 0, \label{Freedman000}
\end{eqnarray}
where $\nabla_\mu\epsilon = \left(\partial_\mu + \frac14
\,{\omega_\mu}^{\nu\rho} \gamma_{\nu\rho}\right)\epsilon$.

The idea of introducing such matrix $\mathbf{W}$ was inspired  by the
earlier work
\cite{LopesCardoso:2001rt,Cardoso:2002ec,Behrndt:2002ee,Cardoso:2002ff,Chamseddine:2001hx}
 on curved BPS domain walls in
(genuine)  five-dimensional $\mathcal{N}=2$ gauged supergravity
\cite{Gunaydin:1984bi,Gunaydin:1985ak,Gunaydin:1999zx,Ceresole:2000jd,Bergshoeff:2004kh},
where the supersymmetry parameters form a pair of symplectic Majorana
spinors, $\epsilon_i$, and the analogue of the superpotential becomes an
$su(2)$-valued    $(2\times 2)$-matrix. This formalism encompasses both
curved and flat domain walls, as the latter are retrieved for a diagonal
$\mathbf W$ matrix.

In this paper, we will refer to \emph{ fake supergravities  as
gravitational theories whose scalar potentials can formally be written in
terms of a superpotential-like matrix as in (\ref{Freedman0a}), such that
the equations of motion for domain walls assume a first order form
compatible with (\ref{Freedman000})}.

The above fake supergravity equations can, in general, capture at most
\emph{part} of the structure of generic $5D$, $\mathcal{N}=2$ gauged
supergravity theories. For one thing, the (typically higher than
one-dimensional)  scalar manifolds of $5D$ vector-, tensor- and
hypermultiplets are subject to a variety of strong geometrical
constraints, none of which are visible in the single-field formalism of
(\ref{Freedman0a})--(\ref{Freedman000}). In practice, the scalar field
geometry requires that the supersymmetry transformation laws and the
scalar potential in $5D$, $\mathcal{N}=2$ gauged supergravity may
generically contain extra terms that do not immediately fit into the
simple fake supergravity set-up.

And yet, this simple set of equations still seems to capture the key
aspects of the BPS equations of true supergravity that are relevant for
domain-wall stability. Therefore, one might  wonder whether there are
perhaps some deeper relations between fake and true supergravity that
only become apparent if one restricts the attention to the effective
dynamics of domain wall solutions. The original motivation of our work
was to clarify precisely this point, i.e., to check to what extent the
fake supergravity of \cite{Freedman:2003ax} is really ``fake'' and under
what circumstances it can   describe true supergravity in five
dimensions. In particular, we wondered under which conditions the single
scalar field language of \cite{Freedman:2003ax} could suffice to encode
the distinct geometrical features of the moduli spaces for scalar fields
belonging to various kinds of  matter multiplets: vector-, tensor- or
hypermultiplets.

{}From the technical point of view, the comparison between fake
supergravity and a specific true supergravity model consists essentially
in the correct identification of the spinor projector and the
superpotential $\mathbf{W}$ from the BPS-equations. One then has to check
whether these identifications are compatible with all true BPS-equations
and whether the scalar potential agrees with (\ref{Freedman0a}). As we
will see, a crucial commutator constraint on the superpotential, which
arises as a consistency condition between specific components of the
``fake'' Killing spinor equations, will serve as an important test in this
analysis.

Interestingly, our attempts to match true and fake supergravity equations
along these lines have driven a fruitful re-investigation of the BPS
equations of $5D$, $\mathcal{N}=2$ supergravity itself and led us to a
number of unexpected and quite general new insights in the context of
curved domain walls. In fact, by completing and clarifying previous
studies
\cite{LopesCardoso:2001rt,Cardoso:2002ec,Behrndt:2002ee,Cardoso:2002ff,Chamseddine:2001hx},
we arrive at a remarkably coherent geometrical picture that illustrates
the different r\^{o}les played by vector-, tensor-, and hypermultiplets.
We find that, on a supersymmetric (curved or flat) domain wall solution,
the BPS equations and the scalar potential can be locally written in the
same form, no matter whether the domain wall is supported by vector- or
by hypermultiplet scalars (we show that tensor scalars cannot play any
r\^{o}le in this set-up). However, despite this formal similarity, the
different geometries governing the vector- and hypermultiplet scalar
manifolds still leave a strong imprint on the solution spaces of these
BPS equations. Indeed, we find that the constraints from the very special
geometry forbid a \emph{curved} BPS-domain wall that is supported solely
by vector multiplet scalars. By contrast, similar constraints are absent
for non-trivial hyper-scalars, either alone or in combination with
running vector scalars. These findings are consistent with the examples
constructed in
\cite{LopesCardoso:2001rt,Cardoso:2002ec,Behrndt:2002ee,Cardoso:2002ff,Chamseddine:2001hx},
which always involved at least one running hyper-scalar.

The above results on curved BPS-domain walls in true $5D$,
$\mathcal{N}=2$ supergravity end up having non-trivial consequences also
for the comparison with fake supergravity, and even suggest the way to
make contact between the two. We show that the BPS-equations and the
scalar potentials of vector and hypermultiplets in true supergravity can
formally be brought to agreement with the analogous expressions in fake
supergravity. However, the impossibility of curved BPS-domain walls
supported solely by vector scalars implies that a \emph{curved}
BPS-domain wall in true supergravity can be described by fake
supergravity only if supported by hyperscalars. It should also be
emphasized that any such coincidence between fake and true supergravity
is, in general, only valid locally along the flow, as it requires some
particular gauge and coordinate choices on the scalar manifolds of
$\mathcal{N}=2$ supergravity that we will precisely identify.

The inverse question, as to whether a given fake supergravity domain wall
can be embedded into true supergravity, involves checking various
constraints required by quaternionic or very special geometry. But at
least for a curved wall, one can immediately rule out that the fake
supergravity scalar sits in a vector multiplet.

Among other things, the analysis presented in this paper might finally
help in deciding whether solutions such as the Janus solution of
\cite{Bak:2003jk}, whose stability was proven in \cite{Freedman:2003ax}
using fake supergravity, can perhaps be embedded into true
five-dimensional supergravity even though it breaks the ten-dimensional
Type IIB supersymmetry it descends from \cite{Karch}.

In spite of the focus on $5D$, ${\cal N} = 2$ supergravity, we stress
that we expect to easily extend our results to ${\cal N} = 2$ theories in
four and six dimensions as they bare the same quaternionic-K{\"a}hler
geometry for hypermultiplets as well as the same action of an $SU(2)$
(sub)group of the $R$--symmetry on the susy spinor.
 It might also be worthwhile to specialize our results to certain interesting
subclasses of theories, such as, e.g.,  the gauged supergravities that
derive from flux compactifications of string theory
\cite{Dall'Agata:2004nw}. Domain wall solutions for this subclass in $4D$
have recently been considered \cite{Mayer:2004sd,LouisVaula}.

The organization of the paper is as follows.
 In Section \ref{fakereview}, we briefly review the fake supergravity
formalism of \cite{Freedman:2003ax}. In Section \ref{realreview}, we then
describe curved BPS-domain walls in $5D$, $\mathcal{N}=2$ supergravity
based on the earlier work
\cite{LopesCardoso:2001rt,Cardoso:2002ec,Behrndt:2002ee,Cardoso:2002ff,Chamseddine:2001hx}.
Section \ref{FakevsReal} constitutes the main part of this paper and is
devoted to the comparison between true and fake supergravity. As a
by-product, we derive the conditions for a BPS-domain wall of true
supergravity to be curved, ruling out the vector scalars as single
supporters. In section \ref{ss:similarities}, we show how the use of
wisely chosen parameterizations  on the scalar manifolds brings the BPS
equations and the scalar potentials for vector and hyper-scalars into an
identical form if one considers these expressions on a BPS-domain wall
solution. We end with some comments in Section \ref{ss:conclusions}.
Appendix \ref{app:SU2choice} gives more details about the $SU(2)$
symmetry and its gauge fixing that is performed in order to obtain
suitable coordinates.

%%%%%%%%%%%%%%%%%%%%%%%%%%%%%%%%%%%%%%%%%%%%%%%%%%%%%%%%%%%%%%%%%%

%Section2: Curved domain walls in fake supergravity

%%%%%%%%%%%%%%%%%%%%%%%%%%%%%%%%%%%%%%%%%%%%%%%%%%%%%%%%%%%%%%

\section{Curved domain walls in fake supergravity}
\label{fakereview}

In this section, we briefly summarize the formalism of fake supergravity
developed in \cite{Freedman:2003ax}, to which we refer the reader  for
further details. Ref.~\cite{Freedman:2003ax} considers scalar gravity
actions of the form
\begin{equation}
S=\int d^{d+1}x \sqrt{-g} \left[ \frac{1}{2\kappa^{2}}R-\frac{1}{2}
\partial_{\mu}\phi\partial^{\mu}\phi -V(\phi) \right], \label{Freedman-1}
\end{equation}
with a scalar potential $V(\phi)$ given by (\ref{Freedman0a}). As
mentioned in the introduction, $\mathbf{W}(\phi)$ is an $su(2)$-valued
$(2\times 2)$-matrix, which implies that quadratic expressions such as
$\mathbf{W}^2$, $(\partial_{\phi}\mathbf{W})^2$ or $\{\mathbf{W},\partial
_\phi \mathbf{W}\}$ are proportional to the unit matrix. This allows one
to write the potential without explicitly taking the trace:
\begin{equation}
V(\phi)\unity=\frac{2(d-1)^2}{\kappa^2} \left[
\frac{1}{\kappa^2}(\partial _\phi \mathbf{W})^2
-\frac{d}{d-1}\mathbf{W}^{2} \right], \label{Freedman0}
\end{equation}
which is the form we will use for our later comparison with true
supergravity.

The matrix $\mathbf{W}$ also enters some ``fake'' Killing spinor
equations for an $SU(2)$-doublet spinor $\epsilon $ as shown in
(\ref{Freedman000}). Using (\ref{curvedmetric}) and assuming that the
scalar $\phi$ depends only on the radial coordinate $r$ (which we choose,
for all $d$,  to be the fifth coordinate $x^5$), (\ref{Freedman000})
reads
\begin{eqnarray}
\left[ \nabla_{m}^{AdS_{d}} +\gamma_{m} \left( \frac{1}{2}
\,U^{\prime}\gamma_{5}
+\mathbf{W} \right) \right] \epsilon &=& 0, \label{Freedman1}\\[2mm]
\left[ \partial_{r}+\gamma_{5}\mathbf{W}\right]
\epsilon &=& 0, \label{Freedman2}\\[2mm]
 \left[ \gamma_{5} \phi^{\prime} - \frac{2(d-1)}{\kappa^2}\partial _\phi
 \mathbf{W}\right] \epsilon &=& 0, \label{Freedman3}
\end{eqnarray}
where $U(r)$ is the warp factor of the metric (\ref{curvedmetric}), and
$\nabla_{m}^{AdS_d}$ denotes the spacetime covariant derivative with only
the spin connection for the $AdS_d$ background metric $g_{mn}(x)$. The
prime means\footnote{For clarity, we changed some notation from
\cite{Freedman:2003ax}.} a derivative with respect to $r$.

It is shown in \cite{Freedman:2003ax} that the system
(\ref{Freedman1})-(\ref{Freedman3}) reproduces the second order field
equations for the warp factor $U(r)$ and the scalar field $\phi(r)$ that
follow from (\ref{Freedman-1}) and (\ref{Freedman0a}) with
\begin{equation}
\gamma^{2}(r) \equiv \left( 1-
\frac{\rme^{-2U(r)}}{2L_{d}^{2}\trace\mathbf{W}^{2}(\phi(r))} \right) =
\frac{\trace\{\mathbf{W},\partial _\phi
\mathbf{W}\}^{2}}{2\trace\mathbf{W}^2\trace(\partial _\phi
\mathbf{W})^{2}}, \label{Freedman12}
\end{equation}
(where $L_d^2=-12/R_{AdS}$ is determined by the scalar curvature of the
AdS space) \emph{provided that} the ``superpotential'' $\mathbf{W}(\phi)$
satisfies the constraint
\begin{equation}
\left[\partial _\phi  \mathbf{W}, \frac{d-1}{\kappa^2}
\partial _\phi\partial _\phi\mathbf{W} +\mathbf{W} \right]=0, \label{Freedman16}
\end{equation}
which is a compatibility condition of (\ref{Freedman2}) and
(\ref{Freedman3}).

Eqs.~(\ref{Freedman12}) and (\ref{Freedman16}) have two important
consequences: (\ref{Freedman12}) implies that a solution where
$\mathbf{W}(\phi)$ is proportional to the first derivative $\partial
_\phi \mathbf{W}(\phi)$ leads to $\gamma (r)=1$, i.e., a flat domain
wall, as $\gamma(r)=1$ implies $L_{d} \rightarrow   \infty $.
Eq.~(\ref{Freedman16}), on the other hand, implies that the
$\phi$-dependence of $\textbf{W}(\phi)$ cannot be arbitrary, but has to
satisfy the commutator constraint  (\ref{Freedman16}). As we will see
later, this consistency condition provides an important test on a true
supergravity theory in order to fit into the framework of ``fake
supergravity''.

Since we are interested in five-dimensional supergravity, from now on we
will specialize the above equations to the case $d=4$ and set
$\kappa^2=1$.

%%%%%%%%%%%%%%%%%%%%%%%%%%%%%%%%%%%%%%%%%%%%%%%%%%%%%

%Section 3: Curved domain walls in $5D$, $\mathcal{N}=2$ gauged supergravity

%%%%%%%%%%%%%%%%%%%%%%%%%%%%%%%%%%%%%%%%%%%%%%%%%%%%%

\section[Curved domain walls in $5D$, $\mathcal{N}=2$ gauged supergravity]
{Curved domain walls in \\ $5D$, $\mathcal{N}=2$ gauged supergravity}
\label{realreview}

In the previous section, we have summarized the fake supergravity
formalism for curved domain walls developed in \cite{Freedman:2003ax}. In
this section, we describe how curved BPS-domain walls arise in true
supergravity. In Section \ref{FakevsReal}, we will then compare the
results for fake and true supergravity and verify to what extent they can
describe the same systems.

\subsection{Five-dimensional, $\mathcal{N}=2$ gauged supergravity}
\label{ss:5dN2SG}

We start by  recalling some of the  most important   features of
five-dimensional, $\mathcal{N}=2$ gauged supergravity theories. Further
technical details can be found in the original references
\cite{Gunaydin:1984bi,Gunaydin:1985ak,Gunaydin:1999zx,Ceresole:2000jd,Bergshoeff:2004kh}.

The matter multiplets that can be coupled to $5D$, $\mathcal{N}=2$
supergravity are vector, tensor and hypermultiplets: the scalar $\phi$ of
the previous section could a priori sit in any of these (or even be a
combination of different types of scalars, as we will see in section
\ref{ss:vectorandhyper}).

The $(n_V+n_T)$ scalar fields of $n_{V}$ vector and  $n_T$ tensor
multiplets parameterize a ``very special'' real  manifold
$\mathcal{M}_{\rm VS}$, i.e., an $\left(n_{V}+n_{T}\right)$--dimensional
hypersurface of an auxiliary $(n_{V}+n_{T}+1)$-dimensional space spanned
by coordinates $h^{\tilde{I}}$  $(\tilde{I}=0,1,\ldots, n_{V}+n_{T}+1)$ :
\begin{equation}
\mathcal{M}_{\rm VS}=\{ h^{\tilde{I}}\in \mathbb{R}^{(n_V+n_T+1)}:
C_{\tilde{I}\tilde{J}\tilde{K}}h^{\tilde{I}}h^{\tilde{J}}h^{\tilde{K}}=1\},
\label{defVS}
\end{equation}
where the constants $C_{\tilde{I}\tilde{J}\tilde{K}}$ appear in a
Chern-Simons-type coupling
 of the Lagrangian.
The embedding coordinates $h^{\tilde{I}}$
 have a natural splitting,
\begin{equation}
h^{\tilde{I}}=(h^I,h^M), \qquad (I=0,1,\ldots,n_{V}), \qquad (M=1,\ldots,
n_{T}),
\end{equation}
where the $h^{I}$ are related to the sub-geometry of the $n_{V}$ vector
multiplets, and the $h^{M}$ refer to the $n_{T}$ tensor multiplets. On
$\mathcal{M}_{\rm VS}$, the $h^{\tilde{I}}$ become functions of the
physical scalar fields, $\varphi^{x}$ $(x=1,\ldots,n_{V}+n_{T})$. The
metric on the very special manifold is determined via the equations
\begin{eqnarray}
  &&g_{xy}=h_x^{\tilde{I}}\,h_{y\tilde{I}},\qquad h_x^{\tilde{I}}\equiv
  -\sqrt{\ft32}\,\partial _xh^{\tilde{I}}, \qquad
  h_{\tilde{I}}\equiv C_{\tilde{I}\tilde{J}\tilde{K}}h^{\tilde{J}}h^{\tilde{K}},
\qquad h_{\tilde{I}x}\equiv \sqrt{\ft32}\,\partial _xh_{\tilde{I}},\nonumber\\
  &&h^{\tilde{I}}h_{\tilde{J}}+h_x^{\tilde{I}}\,g^{xy}\,h_{y\tilde{J}}=
  \delta^{\tilde{I}}_{\tilde{J}}, \qquad h^{\tilde{I}}h_{\tilde{I}}=1,\qquad h^{\tilde{I}}h_{\tilde{I}x}=0.
\label{identVS}
\end{eqnarray}

The scalars  $q^X$ $(X=1,\ldots 4n_{H})$ of  $n_{H}$ hypermultiplets, on
the other hand, take their values in a quaternionic-K{\"a}hler manifold
$\mathcal{M}_{\rm Q}$ \cite{Bagger:1983tt}, i.e., a manifold of real
dimension $4n_H$ with holonomy group contained in $SU(2)\times
USp(2n_H)$. We denote the vielbein on this manifold by $f_X^{iA}$, where
$i=1,2$ and $A=1,\ldots,2n_{H}$ refer to an adapted $SU(2)\times
USp(2n_{H})$ decomposition of the tangent space. The hypercomplex
structure is $(-2)$ times the curvature of the $SU(2)$ part of the
holonomy group\footnote{In fact, the proportionality factor includes the
Planck mass and the metric, which are implicit here.}, denoted as
$\mathcal{R}^{rZX}$ $(r=1,2,3)$, so that the quaternionic identity reads
\begin{equation}
  {\cal R}^r_{XY}{\cal R}^{sYZ}=-\ft14\,\delta ^{rs}\,\delta _X{}^Z
   -\ft12\,\varepsilon ^{rst}\,{\cal R}^t_X{}^Z.
 \label{quaterid}
\end{equation}

Besides these scalar fields, the bosonic sector of the matter multiplets
also contains $n_{T}$ tensor fields $B^M_{\mu\nu}$ $(M=1,\ldots,n_{T})$
from the $n_{T}$ tensor multiplets and
 $n_{V}$  vector fields from the $n_{V}$ vector multiplets.
Including the graviphoton, we thus have a total of $(n_{V}+1)$ vector
fields, $A_{\mu}^{I}$ $(I=0,1,\ldots,n_{V})$, which can be used to gauge
up to $(n_{V}+1)$ isometries of the quaternionic-K{\"a}hler manifold
$\mathcal{M}_{\rm Q}$ %and the very special manifold $\mathcal{M}_{VS}$
(provided such  isometries exist). These symmetries act on the
vector-tensor multiplets by a representation $t_{I\tilde{J}}{}^{\tilde
K}$, where in the pure vector multiplet sector $t_{IJ}{}^K=f_{IJ}{}^K$
are the structure constants, and the other components also satisfy  some
restrictions \cite{Gunaydin:1999zx,Ellis:2001xd,Bergshoeff:2004kh}. The
transformations should leave the defining condition in (\ref{defVS})
invariant, hence
\begin{equation}
  t_{I(\tilde{J}}{}^{\tilde M}C_{\tilde K\tilde L)\tilde M}=0.
 \label{restrgauging}
\end{equation}
The very special K{\"a}hler target space then has Killing vectors
\begin{equation}
  K^{x}_I(\varphi)= -\sqrt{\ft32}t_{I\tilde{J}}{}^{\tilde K}h_{\tilde{K}}^{ x}h^{\tilde J}.
 \label{KillingVSM}
\end{equation}
There may be more Killing vectors, but these are the ones that are gauged
using the gauge vectors in the vector multiplets.

The quaternionic Killing vectors $K_{I}^{X}(q)$ that generate the
isometries on $\mathcal{M}_{\rm Q}$ can be expressed in terms of the
derivatives of $SU(2)$ triplets of Killing prepotentials $P_{I}^{r}(q)$
$(r=1,2,3)$ via
\begin{equation}
D_{X}P^{r}_{I}= \mathcal{R}_{XY}^{r}K^{Y}_{I}, \qquad \Leftrightarrow
\qquad \left\{ \begin{array}{c}
  K_{I}^Y=-\frac{4}{3}\mathcal{R}^{rYX}D_{X}P^{r}_{I} \\ [2mm]
  D_XP_I^r=-\varepsilon ^{rst}{\cal R}^s_{XY}D^YP_I^t,
\end{array}\right.
\label{KillingP}
\end{equation}
where $D_{X}$ denotes the $SU(2)$ covariant derivative, which contains an
$SU(2)$ connection $\omega_{X}^{r}$ with curvature $ \mathcal{R}^r_{XY}$:
\begin{equation}
  D_X P^r= \partial _XP^r+2\,\varepsilon ^{rst} \omega _X^sP^t,\qquad
\mathcal{R}^r_{XY}=2\,\partial _{[X}\omega _{Y]}^r+2\,\varepsilon
^{rst}\omega _X^s\omega _Y^t.
 \label{defSU2cR}
\end{equation}
The prepotentials satisfy the constraint
\begin{equation}\label{constraint}
\frac{1}{2}\mathcal{R}_{XY}^{r}K_{I}^{X}K_{J}^{Y}-
\varepsilon^{rst}P_{I}^{s}P_{J}^{t} +\frac{1}{2}f_{IJ}{}^{K}P_{K}^{r}=0,
\end{equation}
where $f_{IJ}{}^{K}$ are the structure constants of the gauge group.

In the following, we will frequently switch between the above
vector notation
for $SU(2)$-valued quantities such as $P_{I}^{r}$, and  the usual
   $(2\times 2)$   matrix notation,
\begin{equation}
\mathbf{P}_I= \left(   P_{Ii}{}^j\right) ,\qquad
%  \equiv\rmi \vec \sigma_i{}^j\cdot\vec P_{I}
P_{Ii}{}^j\equiv \rmi\,\sigma_{ri}{}^jP_{I}^{r}  .
 \label{matrixVector}
\end{equation}
As in \cite{Freedman:2003ax}, boldface expressions such as $\mathbf{P}_I$
then refer to the $(2\times 2)$-matrices with the indices $i,j$
suppressed.

An important difference in geometrical significance between the very
special Killing vectors $K_{I}^{x}(\varphi)$ in  (\ref{KillingVSM})  and
the quaternionic ones $K_{I}^{X}(q)$  in (\ref{KillingP}), is that the
former do not arise as derivatives of Killing prepotentials, because
there is no natural symplectic structure on the real manifold
$\mathcal{M}_{\rm VS}$ that could define a moment map. This feature will
also play a r\^{o}le in the comparison with fake
supergravity.\footnote{The moment maps are related to the fact that the
isometries should preserve complex structures. Therefore, they are absent
in the real manifold. In 4 dimensions, the scalar manifold of the vector
multiplets does have a complex structure. Hence, in that case this sector
would also have a moment map structure~\cite{D'Auria:1991fj}. This
suggests that in four dimensions the same comparison may go along
different lines.}

Turning on only the metric and the scalars, the general Lagrangian of
such a gauged supergravity  theory
is
\begin{equation}
e^{-1}\mathcal{L}=-\frac{1}{2}R-\frac{1}{2}g_{xy}\partial_{\mu}\varphi^{x}
\partial^{\mu}\varphi^{y}-\frac{1}{2}g_{XY}\partial_{\mu}q^{X}
\partial^{\mu}q^{Y}-g^2\mathcal{V}(\varphi,q),
\end{equation}
whereas the supersymmetry transformation laws of the fermions are given by
\begin{eqnarray}
\delta \psi_{\mu i}&=& {\nabla}_\mu\epsilon_i-  \omega_{\mu i}{}^{j}
\epsilon_{j}- \frac{\rmi}{\sqrt{6}}\,
g\,\gamma_{\mu}P_{i}^{\,\, j}\epsilon_{j},\label{gravitino}\\
\delta \lambda_{i}^{x}&=&
-\frac{\rmi}{2}\gamma^{\mu}(\partial_{\mu}\varphi^{x})\epsilon_{i}
-g\,P_{i}{}^{jx}\epsilon_{j}+g\,\mathcal{T}^x\epsilon_{i},\label{gaugino}\\
\delta
\zeta^{A}&=&\frac{\rmi}{2}f_{X}^{iA}\gamma^{\mu}(\partial_{\mu}q^{X})
\epsilon_{i} - g\, \mathcal{N}^{iA}\epsilon_{i}.\label{hyperino}
\end{eqnarray}
Here, $\psi_{\mu}^{i}$, $\lambda_{i}^{x}$, $\zeta^{A}$ are the gravitini,
gaugini (tensorini) and hyperini, respectively,
 $g$ denotes the gauge coupling, the $SU(2)$ connection $\upomega_{\mu}$ is
defined as $\omega_{\mu i}{}^{j}= (\partial_{\mu}q^{X}) \omega_{X
i}{}^{j} $, and
\begin{eqnarray}
P^{r}&=&h^{I}(\varphi)P_{I}^{r}(q), \label{Pdefinition}  \\
P^{r}_{x}&=&-\sqrt{\frac{3}{2}}\partial_{x}P^r=h_{x}^{I}P_{I}^{r},\qquad P^{rx}=g^{xy}P^r_y,\\
\mathcal{N}^{iA}&=&\frac{\sqrt{6}}{4}f_X^{iA}(q)h^{I}(\varphi)K_{I}^{X}(q),\label{hshift}\\
\mathcal{T}^{x}&=&\frac{\sqrt{6}}{4}h^{I}(\varphi)K_{I}^{x}(\varphi)
%=-\frac34t_{I\tilde J}{}^{\tilde K}h^Ih^{\tilde J}h^x_{\tilde K}
\label{tshift}.
\end{eqnarray}
As a general fact in supergravity, the potential is given by the sum of
``squares of the fermionic shifts'' (the scalar expressions in the above
transformations of the fermions):
\begin{equation}
\mathcal{V}=-4P^{r}P^{r} +2P_{x}^{r}P_{y}^{r}g^{xy}+2\mathcal{N}^{iA}
\mathcal{N}^{jB}\varepsilon _{ij}C_{AB}+2\mathcal{T}^x
\mathcal{T}^yg_{xy}, \label{scalarpotential}
\end{equation}
where $C_{AB}$ is the (antisymmetric) symplectic metric of $USp(2n_{H})$.

A first glance at (\ref{gravitino}) indicates that, leaving out for the
moment the $SU(2)$ connection term $\upomega_{\mu}$, the superpotential
matrix $\mathbf{W}$ has to be related to $\mathbf{P}$.

Using the explicit form of the Killing vector, (\ref{KillingVSM}), in
(\ref{tshift}), one finds that this expression vanishes if the
transformation matrix $t$ involves only vector multiplets. This is clear
because then $t_{IJ}{}^K=f_{IJ}{}^K$, hence antisymmetric. Therefore, the
shift $\mathcal{T}^{x}$ in the above expressions is non-vanishing only if
there are charged tensor multiplets in the theory \footnote{In five
dimensions, tensor multiplets that are not charged under some gauge group
are equivalent to vector multiplets. We always assume that all uncharged
tensor multiplets are converted to vector multiplets.}. Since
$\mathcal{T}^{x}$ appears in (\ref{gaugino}) with the unit matrix in
$su(2)$ space, it must vanish on a BPS-domain wall solution for
compatibility with the spinor projector (see \cite[footnote
8]{Ceresole:2001wi} and \cite{Ceresole:2001zf}). Furthermore, unlike the
shifts $P_{x}^{r}$ and $\mathcal{N}^{iA}$, $\mathcal{T}^{x}$ is a purely
``D-type'' term, in the sense that it is completely unrelated to
derivatives of the matrix $\mathbf{P}$. Therefore, it can never fit the
pattern (\ref{Freedman000}) of the fake supergravity transformations.
Thus, neither for BPS-domain walls in $5D$, $\mathcal{N}=2$ supergravity
nor for domain walls in fake supergravity, can non-trivial tensor
multiplets play an important r\^{o}le, and we can limit our remaining
discussion to the case  $n_T=0$, i.e., to supergravity coupled to vector
and/or hypermultiplets only. This also means that the index $\tilde{I}$
simply becomes the index $I$ in all previous equations, and the index $M$
disappears.

Using (\ref{KillingP}) and the quaternionic identity (\ref{quaterid}),
the scalar potential for vector and hypermultiplets can be written in a
form that is somewhat similar to (\ref{Freedman0}),
\begin{equation}
\mathcal{V}\unity =4 \mathbf{P}^2 -3
(\partial_{x}\mathbf{P})(\partial^{x}\mathbf{P})
-(D_X\mathbf{P})(D^{X}\mathbf{P})\label{VPP2}.
\end{equation}

In Sections  \ref{FakevsReal}  and \ref{ss:similarities}, we will
elaborate further  on the similarities and differences between the true
and fake supergravity potentials, as well as on how to remove the
asymmetry between the hypermultiplet and the vector multiplet sector in
these expressions.

%%%%%%%%%%%%%%%%%%%%%%%%%%%%%%%%%%%%%%%%%%%%%%%%%%%%%%%%%%%%%%%%%%%

\subsection{Curved BPS-domain walls in supergravity}

We can now take a closer look at $1/2$ supersymmetric, curved domain wall
solutions   of the above gauged supergravity theories. The careful
investigation of this subject was pioneered in \cite{LopesCardoso:2001rt}
and further developed in  refs.
\cite{Cardoso:2002ec,Behrndt:2002ee,Cardoso:2002ff,Chamseddine:2001hx},
where also some examples were given. Here we mainly review this
construction, although in a different language and deriving a new
important constraint.

In a curved domain wall background of the form (\ref{curvedmetric}), when
the scalar fields only depend on the radial coordinate, the vanishing of
the supersymmetry variations (\ref{gravitino})-(\ref{hyperino}) implies
\begin{eqnarray}
  \left[
\nabla_{m}^{AdS_4} +\gamma_{m} \left( \frac{1}{2} U^{\prime}\gamma_{5}
-\frac{ig}{\sqrt{6}}\mathbf{P} \right) \right] \epsilon &=&0,
\label{realgravitino1}\\
\left[D_{r}   + \gamma_{5} \left( -\frac{\rmi g}{\sqrt{6}} \mathbf{P}
\right)
\right] \epsilon &=&0,  \label{realgravitino2} \\
\left[  \gamma_{5}\varphi^{x\prime} + \rmi
g\,\sqrt{6}\,g^{xy}\partial_{y} \mathbf{P}
\right] \epsilon &=&0, \label{realgaugino}\\
 f_{X}^{iA}\left[\gamma_{5}q^{X\prime}- \rmi g \sqrt{\frac{8}{3}}R^{rXY}D_Y P^r \right]
\epsilon_{i}   &=& 0, \label{realhyperino}
\end{eqnarray}
where
\begin{equation}
D_r\epsilon_{i}\equiv \partial_{r}\epsilon_{i}
 -q^{X\prime}\omega_{Xi}{}^{j} \epsilon_{j} \label{Dr}
\end{equation}
has been introduced.

These equations have a structure similar to (\ref{Freedman000})
\begin{eqnarray}
\delta \psi_{\mu i}&=&\left( {\nabla}_{\mu}\delta _i{}^j-  \omega_{\mu
i}{}^{j}  - \frac{\rmi}{\sqrt{6}} \,g\,\gamma_{\mu}P_i{}^j\right)\epsilon_{j}=0,\label{gravitino1}\\
 \delta \lambda_{i}^{x}&=&
\mathcal{O}^x{}_i{}^j\epsilon_{j}=0, \label{gaugino1}\\
f^X_{iA}\delta\zeta^{A}&=& \mathcal{O}^X{}_i{}^j \epsilon_{j}=0
,\label{hyperino1}
\end{eqnarray}
and it is useful to split the operators $\mathcal{O}_i{}^j$ into
$\mathcal{O}_i{}^j=\mathcal{O}^0 \delta_i{}^j+\rmi \mathcal{O}^{r}
\sigma_{ri}{}^j$, whose components in each case can be read off directly
from the explicit formulae.

In order to construct solutions to the system
(\ref{realgravitino1})--(\ref{realhyperino}), one usually splits these
equations into a projection condition on the supersymmetry parameter
$\epsilon$ and a system of first order differential equations involving
only the scalars and the warp factor. To do so, extending the ideas of
\cite{Ceresole:2001wi} to curved domain walls, one chooses a projector on
the supersymmetry parameter of the form
\begin{equation}
 \ifiTheta
  \epsilon _i=-\gamma _5\Theta _i{}^j\epsilon _j    \Leftrightarrow
\left(\unity +\gamma_{5}\mathbf{\Theta}\right)\epsilon =0,
  \else
  \epsilon _i=-\rmi\gamma _5\Theta _i{}^j\epsilon _j    \Leftrightarrow
\left(\unity +\rmi\gamma_{5}\mathbf{\Theta}\right)\epsilon =0,
  \fi
 \label{projepsilon}
\end{equation}
where $\mathbf{\Theta} ^2=\ifiTheta \else -\fi\unity   \Leftrightarrow
\Theta^r \Theta^r =\ifiTheta -\else \fi 1$.

The explicit form of $\mathbf \Theta$ can be determined by the solution
of the supersymmetry transformations on the matter fields, $\delta
\lambda_i^x$ and $\delta \zeta^A$. More precisely,   using
(\ref{projepsilon}), the gaugino transformation (\ref{realgaugino})
implies the BPS-equation
\begin{equation}
g_{yx}\,\varphi^{x\prime}\mathbf{ \Theta} =
\ifiTheta\rmi\else\fi\sqrt{6}\,g\,\partial_{y}\mathbf{P}.
\label{vectorBPS1}
\end{equation}
Note that we can omit $\epsilon $ as the only projection on the Killing
spinor involves $\gamma _5$. Hence
\cite{LopesCardoso:2001rt,Cardoso:2002ec}
\begin{equation}
\mathbf{\Theta}=\ifiTheta\rmi\else\fi g\,\sqrt{6}\,
\frac{\varphi^{x\prime}\partial_{x}\mathbf{P}}{\varphi^{y\prime}
\varphi^{z\prime}g_{yz}}.\label{Thetavector}
\end{equation}

The hyperino transformation (\ref{realhyperino}), on the other hand,
implies, after contraction with $f_{YjA}$ and
 a decomposition into the trace and the traceless  part
(see \cite{Ceresole:2001wi} for the flat domain wall  analogue),
\begin{eqnarray}
\ifiTheta\rmi\,\else\fi
g_{YX}q^{X\prime}\mathbf{\Theta}+\ifiTheta\rmi\else\fi
q^{X\prime}[\mathbf{R}_{YX},\mathbf{\Theta}] \ifiTheta +\else -\fi
\sqrt{6}\,g\, D_{Y}\mathbf{P} &=&0,
\label{BPS1}\\
\sqrt{6}\,g\,K_{Y}\ifiTheta+ 2\rmi\else-2\fi \, q^{X\prime} \{
\mathbf{R}_{YX},\mathbf{\Theta} \}&=&0 . \label{BPS2}
\end{eqnarray}
These two equations are equivalent and can be converted into one another
via contractions with the $SU(2)$ curvature. Another interesting and
compact version\footnote{The relation (\ref{usefulform}) can be directly
derived by applying the general analysis of the  hyperino equation in
\cite{Celi:2003qk,Cacciatori:2004zz}.} of the hyperino equation can be
obtained by anticommuting (\ref{BPS1}) with $\mathbf{\Theta}$:
\begin{equation}
g_{XY}q^{X\prime}=\ifiTheta-\rmi\else\fi g\,\sqrt{6} \,\Theta^r D_{Y}
P^r. \label{usefulform}
\end{equation}
Comparing (\ref{BPS1}) with      (\ref{vectorBPS1}), we notice again an
obvious asymmetry   between the vector and hypermultiplets. Contraction
of (\ref{BPS1})  with $q^{X\prime}$ finally  yields the expression
\cite{Behrndt:2002ee}
\begin{equation}
\mathbf{\Theta}=\ifiTheta\rmi\else\fi g\,\sqrt{6}\,\frac{q^{X\prime}D_{X}
\mathbf{P}}{q^{Y\prime} q^{Z\prime} g_{YZ}}\label{Thetahyper}.
\end{equation}
When both vector multiplet scalars and hyper-scalars are non-trivial,
consistency of (\ref{Thetahyper}) and (\ref{Thetavector}) requires
\begin{equation}\label{consistency}
\frac{q^{X\prime}D_{X}\mathbf{P}}{q^{Y\prime}q^{Z\prime}g_{YZ}}=
\frac{\varphi^{x\prime}\partial_{x}\mathbf{P}}{\varphi^{y\prime}
\varphi^{z\prime}g_{yz}}.
\end{equation}

On the other hand, (\ref{Thetavector}) and (\ref{Thetahyper}) also imply
that $\mathbf{\Theta}$ is proportional to $D_r\mathbf{P}$:
\begin{equation}
  D_r\mathbf{P}\equiv \varphi ^{\prime x}\partial _x\mathbf{P}+q^{\prime
  X}D_X\mathbf{P}= \ifiTheta-\frac \rmi{\sqrt 6 g}\else \frac 1{\sqrt 6 g}\fi
   g_{\Lambda\Sigma}\phi^{\Lambda\prime}\phi^{\Sigma\prime}
  \mathbf{\Theta},
 \label{DrPproptoTheta}
\end{equation}
where $\phi^{\Lambda}=\{\varphi^{x}, q^{X}\}$.

There is, however, one further integrability constraint that was not
noticed before. It follows from consistency  between (\ref{projepsilon})
and (\ref{realgravitino2}). This relation will play a r\^{o}le analogous
to the consistency condition (\ref{Freedman16}) of  the fake supergravity
framework. To this end, consider
 \ifiTheta
    \begin{eqnarray}
    0 &\stackrel{(\ref{projepsilon})}{=}&
     \mathbf{\Theta}\gamma_{5}D_{r}[\epsilon+\gamma_{5}\mathbf{\Theta}\epsilon]\nonumber\\
     &\stackrel{(\ref{realgravitino2})}{=}& \frac{\rmi
     g}{\sqrt{6}}\,[\mathbf{\Theta}\mathbf{P}\epsilon
     +\gamma_{5}\mathbf{P}\epsilon]+\mathbf{\Theta}(D_r \mathbf{\Theta})
     \epsilon\nonumber\\
      &\stackrel{(\ref{projepsilon})}{=}& \left[
     \frac{\rmi g}{\sqrt{6}}
     \, [\mathbf{\Theta},\mathbf{P}]+\mathbf{\Theta}(D_{r}\mathbf{\Theta})
     \right] \epsilon.
    \end{eqnarray}
 \else
    \begin{eqnarray}
     0 &\stackrel{(\ref{projepsilon})}{=}&
     -\rmi\mathbf{\Theta}\gamma_{5}D_{r}[\epsilon+\rmi\gamma_{5}\mathbf{\Theta}\epsilon]
     \nonumber\\
     &\stackrel{(\ref{realgravitino2})}{=}& \frac{
     g}{\sqrt{6}}\,[\mathbf{\Theta}\mathbf{P}\epsilon
     -\rmi\gamma_{5}\mathbf{P}\epsilon]+\mathbf{\Theta}(D_r \mathbf{\Theta})
     \epsilon\nonumber\\
      &\stackrel{(\ref{projepsilon})}{=}& \left[
     \frac{g}{\sqrt{6}}
    \, [\mathbf{\Theta},\mathbf{P}]+\mathbf{\Theta}(D_{r}\mathbf{\Theta})
    \right] \epsilon.
    \end{eqnarray}
 \fi
 Remembering  $D_{r}(\mathbf{\Theta}^{2})=0$, the second term in the last equation can also be written as $\frac{1}{2}[\mathbf{\Theta},D_{r}\mathbf{\Theta}]$,
and one finally obtains the consistency condition\footnote{An analogous
equation was independently derived by Klaus Behrndt and Mirjam Cveti\v{c}
(private communication).}
\begin{equation}
  \left[\mathbf{\Theta }\,,\,D_r \mathbf{\Theta}  +\ifiTheta\rmi\else\fi\sqrt{\frac23}g\,
\mathbf{P}\right]=0.
 \label{commutator1}
\end{equation}
Using (\ref{DrPproptoTheta}) this leads to the commutator relation
\begin{equation}
  \left[ D_r\mathbf{P},D_rD_r\mathbf{P}
  +\frac 13 g_{\Lambda\Sigma}\phi^{\Lambda\prime}\phi^{\Sigma\prime}\mathbf{P}\right]
  =0,
 \label{commutatorD}
\end{equation}
which is the supergravity version of the equation (\ref{Freedman16}) in
fake supergravity. The differences between (\ref{commutatorD}) and
(\ref{Freedman16}) are the fact that (\ref{commutatorD}) holds with
derivatives taken in the (typically higher-dimensional) space of all the
scalar fields in the supergravity theory and the consequent appearance of
covariant derivatives.

It should be noted that the conditions summarized in this section are
just necessary conditions in order to obtain supersymmetric domain wall
solutions in five-dimensional supergravity. To verify whether they are
also sufficient, one has to further check the equations of motion. These
are identically satisfied for $\gamma = 1$, i.e. for flat domain walls,
but may give additional constraints in the case of curved ones.

%%%%%%%%%%%%%%%%%%%%%%%%%%%%%%%%%%%%%%%%%%%%%%%%%%%%%%%%%%%%%%%%%%%%%%%%%%%%%

%   Section 4: Fake vs. Real supergravity

%%%%%%%%%%%%%%%%%%%%%%%%%%%%%%%%%%%%%%%%%%%%%%%%%%%%%%%%%%%%%%%%%%%%%%%%

\section{True vs. fake supergravity}
\label{FakevsReal}

In this section, we want to find out whether there are supersymmetric
\emph{curved}  domain walls in true supergravity that  can also be
described within the simpler framework of fake supergravity. The most
obvious obstacle for such a comparison  is the number of scalar fields
within these two setups. While the fake supergravity formalism of
\cite{Freedman:2003ax} was developed in detail for only one scalar field
\footnote{A generalization to more scalar fields with particular types of
scalar potentials is briefly discussed in \cite{Freedman:2003ax}.},
$\phi$, %and contained no information about the
%geometry of the corresponding moduli space,
a generic true %real
supergravity theory contains $(n_{V}+4n_{H})$ scalar fields (we have
already discarded the possibility of tensor multiplet scalars)
$\phi^{\Lambda}=(\varphi^{x},q^{X}) $ which exceeds one unless there is
precisely one vector and no hypermultiplet. Comparing these two setups is
thus only possible if the ``superfluous'' scalars in true supergravity
can somehow be ``deactivated''. As we will now describe, for any given
domain wall solution, one can, in principle, always choose  local,
adapted coordinate systems on the scalar manifolds of true supergravity
such that there is at most  one $r$-dependent  scalar field on
$\mathcal{M}_{VS}$ and $\mathcal{M}_{Q}$,  and all other scalar fields
can effectively be removed from the equations that describe the domain
wall. It is in these adapted coordinates that we will be able to
investigate the question as to whether fake supergravity also describes
some true supergravity systems, as they allow to reduce the discussion
essentially to one single scalar.

The usefulness of adapted coordinates goes beyond the comparison of fake
and true supergravity. Indeed, as another interesting application, we
will show that the differences between vector and hypermultiplets can be
switched off along the scalar flow curve when adapted coordinates are
used. With this rewriting, we are then able to sharpen the conditions for
curved domain walls to exist in true supergravity. In particular, we show
that supersymmetric  domain walls that are supported by vector scalars
only, have to be flat.

%%%%%%%%%%%%%%%%%%%%%%%%%%%%%%%%%%%%%%%%%%%%%%%%%%%%%%%%%%%%

\subsection{Running vector scalars}
\label{ss:runvectors}

To begin with, let us investigate the possibility that the scalar field
$\phi$ of the fake supergravity equations
(\ref{Freedman-1})-(\ref{Freedman3}) belongs to a vector multiplet. For
this to be possible, the curved domain wall obviously has to be supported
by vector scalars only, i.e., any hyper-scalars (if present)
 have to stay constant along the flow:
\begin{equation}
q^{X\prime}=0.
\end{equation}
For this condition to preserve supersymmetry, the hyperino BPS-equation
  (\ref{BPS1}) implies
that
\begin{equation} D_{X}\mathbf{P}=0 \label{DXP0}
\end{equation}
along the flow.

The constancy of the hyper-scalars also means that the $SU(2)$ connection
$q^{X\prime}\omega_{Xi}{}^{j}$ vanishes along the flow, and the $SU(2)$
covariant derivative $D_{r}$ defined in (\ref{Dr}) degenerates to an
ordinary derivative, $D_{r}=\partial_{r}$.

It is easy to see that the gravitino BPS-equations
(\ref{realgravitino1})-(\ref{realgravitino2}) then become equivalent to
the fake supersymmetry transformations
(\ref{Freedman1})-(\ref{Freedman2}), provided we identify
\begin{equation}
\mathbf{W} = -\frac{\rmi g}{\sqrt{6}} \mathbf{P}.
 \label{WP}
\end{equation}

In order to bring the   gaugino BPS-equation  (\ref{vectorBPS1}) into the
form of fake supergravity, one has to get rid of all but one scalar
field, which is done by going to a particular, ``adapted'' coordinate
system on the scalar manifold   $\mathcal{M}_{\rm VS}$. To this end,
recall  that, in general, any   domain wall solution defines a curve
$\phi^{\Lambda}(r)$ on the scalar manifold $\mathcal{M}=\mathcal{M}_{\rm
VS}\times \mathcal{M}_{Q}$. In this subsection, we only consider curves
that are trivial on the quaternionic manifold and therefore lie entirely
on $\mathcal{M}_{\rm VS}$. As the coordinates $\varphi^{x}$ on
$\mathcal{M}_{\rm VS}$ can be chosen at will, we can take a basis for
these scalars such that only one scalar of the vector multiplets  is
$r$-dependent and all  the others are constant \footnote{In practice,
this might be a very inconvenient choice to work with, and it might also
obscure the one-to-one correspondence between particular scalar fields
and certain gauge theory operators in an AdS/CFT context. It is also
clear that, by construction, these adapted coordinates are different for
different flow solutions and that one might have to use several adapted
coordinate patches   to cover an entire flow curve, as such a curve might
have self-intersections. For our formal arguments, however, this
coordinate choice turns out to be  convenient and sufficient.}. We call
this single $r$-dependent scalar field $\varphi$. The other scalars,
which we call $\varphi^{\hat{x}}$, can then be chosen to be orthogonal to
the flow curve (at least on the flow curve itself). Although this last
assumption is not strictly necessary to derive most of the results we
present here, it is always possible in some patch and therefore we employ
it for the sake of simplicity. Along this flow curve, the scalar field
metric $g_{xy}$ then takes the form
\begin{equation}
g_{xy}= \left( \begin{array}{cc}
g_{\varphi\varphi} & 0 \\
0 & g_{\hat{x}\hat{y}}
\end{array} \right). \label{adaptedmetric}
\end{equation}
Note that in these coordinates, the vanishing of the gaugino
transformation (\ref{realgaugino}) for the constant scalars
$\varphi^{\hat{x}}$ implies
\begin{equation}
\partial_{\hat{x}}\mathbf{P}=0 \label{hatP}
\end{equation}
on the flow curve. The  $\varphi $-component of (\ref{realgaugino}), on
the other hand,   is now of the form (\ref{Freedman3}) given in  fake
supergravity. Given the identification (\ref{WP}) and identifying the
scalar of fake supergravity $\phi $ with $\varphi$, it is then easy to
see that
\begin{itemize}
  \item The supersymmetry transformations (\ref{realgravitino1})-(\ref{realgaugino})
are of the same form as the  fake supersymmetry transformation
(\ref{Freedman1})-(\ref{Freedman3});
  \item The consistency condition (\ref{commutatorD}) reduces to
\begin{equation}
 \left[
   \partial_{\varphi}\mathbf{P}, \partial^{2}_{\varphi}\mathbf{P}
+\frac{1}{3}g_{\varphi \varphi } \mathbf{P}     \right] =0 ,
\label{comp2}
\end{equation}
which is equivalent to the required compatibility condition
(\ref{Freedman16}), if one normalizes $g_{\varphi \varphi }=1$ by an
appropriate rescaling of $\varphi $;
  \item Upon this normalization, also
  the Lagrangian and the scalar potentials agree, see (\ref{Freedman0})
  versus (\ref{VPP2}) with (\ref{DXP0}) and (\ref{hatP}).
\end{itemize}
Thus, at first sight,   the case of running vector scalars in genuine
$5D$, $\mathcal{N}=2$ gauged supergravity automatically seems to fall
into the generalized fake supergravity framework of
\cite{Freedman:2003ax}. We will now show that this conclusion is
\emph{wrong} when the domain wall is supposed to be \emph{curved}.

The crucial point is the consistency condition (\ref{comp2}). This
equation arose  as a consistency condition between the gravitino and the
gaugino supersymmetry variations in our desired curved domain wall
background. As it stands, it is a constraint on the possible field
dependence of the matrix $\mathbf{P}$. Supergravity, on the other hand,
already constrains this matrix   function independently of any desired
background solution. As we will  now show,   these supergravity
constraints are so strong that (\ref{comp2})
  cannot be satisfied non-trivially in a \emph{curved} domain wall background
 for scalar fields that sit in
vector multiplets. This rules out the possibility that the scalar field
$\phi$ of fake supergravity with a genuine matrix superpotential can be a
scalar sitting in a vector multiplet. Moreover, it shows that  a
BPS-domain wall  in true supergravity that is supported by vector scalars
only can at most be flat.

In order to see this, we observe that the derivatives of $\mathbf{P}$
with respect to the vector scalars $\varphi^{x}$ are, from its definition
(\ref{Pdefinition}), determined by the derivatives of $h^{I}$:
\begin{equation}
  \partial_x \mathbf{P} = - \sqrt{\frac 23} \mathbf{P}_I h^I_x,\qquad
  \partial_y\partial_x \mathbf{P} = - \sqrt{\frac 23} \mathbf{P}_I \partial_y h^I_x.
 \label{Pderivatives}
\end{equation}
{}From the second line of (\ref{identVS}), one obtains
\begin{eqnarray}
  \partial_y h_x^I &=& \partial_y h_x^J \left( h_J h^I + h_J^z h^I_z\right) =
 - \sqrt{\ft23}h_x^J h_{Jy}h^I+ \partial_y h_x^J h_J^z h^I_z,\nonumber\\
  \partial_y \partial_x  \mathbf{P} &=& \ft 23 g_{xy}\mathbf{P}+
  (\partial_y h_x^J h_J^z) \partial_z  \mathbf{P}.
 \label{ddP}
\end{eqnarray}
In adapted coordinates $(\varphi(r),\varphi^{\hat{x}})$ with (\ref{hatP})
and metric (\ref{adaptedmetric}), this becomes
\begin{equation}
\partial^{2}_{\varphi}\mathbf{P}=\frac{2}{3}g_{\varphi \varphi } \mathbf{P} +
\textrm{ (terms proportional to } \partial_{\varphi}\mathbf{P}\textrm{)}
\label{veryspecialidentity}
\end{equation}
and hence
\begin{equation}
\left[ \partial_{\varphi}\mathbf{P}, \partial^{2}_{\varphi}\mathbf{P}-
\frac{2}{3}g_{\varphi \varphi }\mathbf{P} \right]
=0,\label{trueconsistency}
\end{equation}
which differs from the desired relation (\ref{comp2}). The only way,
(\ref{trueconsistency}) and (\ref{comp2}) could be reconciled would be to
demand that, along the flow curve, $[\partial_{\varphi}
\mathbf{P},\mathbf{P}]=0  $, or equivalently, $  \partial _\varphi
\mathbf{W}=f(\varphi) \mathbf{W} $, with some function $f(\varphi)$,
which, however,  would then imply $\gamma(r)=1$ via (\ref{Freedman12}),
i.e. a flat domain wall\footnote{Because of
$\partial_{\hat{x}}\mathbf{P}=0$ in adapted coordinates, this is nothing
but the condition $\partial_{x}Q^{s}=0$  for flat domain walls
 of \cite{Ceresole:2001wi},
where $Q^s$ denotes the phase of $P^s$ [cf. (\ref{splitPWQ})].}. Thus, we
conclude that a BPS domain wall in $5D$, $\mathcal{N}=2$ supergravity
that is supported by vector scalars only, can at most  be flat.
Therefore, the curved domain walls of \cite{Freedman:2003ax} cannot be
the ones described by true supergravity where only the scalars of vector
multiplets are running.

However, for the flat domain walls we find indeed agreement as
$[\partial_{\varphi}\mathbf{P}, \mathbf{P}]=0$ is always satisfied. This
can be proven as follows. We assume no running hyper-scalars (or the
situation without hyper-scalars), i.e. the $q^{X}$ sit at a critical
point $q^{X}_{0}$, such that due to (\ref{BPS2}),
\begin{equation}\label{quat}
h^{I}(\varphi)K_{I}^{X}(q_{0})=0.
\end{equation}
Then (\ref{constraint}) contracted with $h^{I}_{x}$ and $h^{J}$ implies
\begin{equation}\label{trivial}
[\partial _x\mathbf{P},\mathbf{P}]=0.
\end{equation}
The proof is obvious in the case of an Abelian gauge group, but holds
also in the non-Abelian case, making use of (\ref{identVS}), the
invariance requirement on the coefficients $C_{IJK}$ leading to
$f_{IJ}{}^Kh_Kh^J=0$ \cite{Gunaydin:1985ak}, and (\ref{hatP}).
This reconciles clearly (\ref{comp2}) with (\ref{trueconsistency}).

%%%%%%%%%%%%%%%%%%%%%%%%%%%%%%%%%%%%%%%%%%%%%%%%%%%%%%%%%%

%   Section 4.2: Running hyper-scalars

%%%%%%%%%%%%%%%%%%%%%%%%%%%%%%%%%%%%%%%%%%%%%%%%%%%%%%%%%%

\subsection{Running hyper-scalars} \label{ss:runhyperscalars}

In this section, we will consider curved BPS-domain walls that are
supported by hyper-scalars only, i.e., we will assume that all potential
vector scalars are constant:
\begin{equation}
\varphi^{x\prime}=0.
\end{equation}
The possibility that both vector scalars and hyper-scalars are running
will be considered in section \ref{ss:vectorandhyper}. The gaugino
BPS-equations (\ref{realgaugino}) now imply
\begin{equation}
\partial_{x}\mathbf{P}=0
\end{equation}
for consistency.

We now turn to the other BPS equations. If we again make the
identification (\ref{WP}), it is easy to  see that, modulo the
$SU(2)$-connection $q^{X\prime} \omega_{Xi}{}^{j}$, the gravitino
BPS-equations (\ref{realgravitino1})-(\ref{realgravitino2}) are again of
the same  form as the corresponding equations
(\ref{Freedman1})-(\ref{Freedman2}) of fake supergravity. We are thus led
to the question as to whether the $SU(2)$-connection can be gauge fixed
in such a way as to reproduce exactly  equations
(\ref{Freedman1})-(\ref{Freedman2}). To answer this question, note that
we only need the vanishing of this $SU(2)$ connection in one direction
(the one of $q^{X\prime}$). Thus, if one can achieve
\begin{equation}
SU(2)\mbox{ gauge choice:}\qquad   q^{X\prime} \omega_X^r=0,
 \label{SU2gauge}
\end{equation}
the gravitino equations in fake and true supergravity with running
hyper-scalars, locally,   agree. However, on a sufficiently short segment
of the flow curve, this gauge can always be achieved by simply taking the
relevant  gauge transformation equal to the inverse of the Wilson line of
the original $SU(2)$-connection along that curve segment. This is further
explained and formalized in Appendix \ref{app:SU2choice}.

Before we proceed, we would like to comment on the validity of the local
$SU(2)$-symmetry that underlies this gauge choice. Geometrically, the
local $SU(2)\times USp(2n_{H})$ invariance is the part of the naive
$SO(4n_{H})$ tangent space group of the target manifold $\mathcal{M}_{Q}$
that survives the supersymmetric coupling to the fermions. As such, this
local composite invariance should not interfere with the gaugings of
isometries of the target space metric $g_{XY}$, as the latter is
manifestly  invariant under the $SU(2)\times USp(2n_{H})$
reparametrizations of the quaternionic vielbeins $f_{X}^{iA}$. And
indeed, as can be read off explicitly from the expressions in
\cite{Ceresole:2000jd,Bergshoeff:2004kh}, the gauged Lagrangian and
supersymmetry transformations are still manifestly invariant and
covariant, respectively, with respect to $SU(2)$ (and $USp(2n_{H})$). The
BPS-equations for domain wall solutions, in which the vector fields and
fermions are set to zero, also inherit this $SU(2)$ covariance, i.e., any
BPS-domain wall is part of an $SU(2)$ orbit of gauge equivalent
solutions, and we are free to partially fix that gauge symmetry in the
way we do above and in Appendix \ref{app:SU2choice}.

Such a gauge choice thus restricts the form of the quaternionic
vielbeins, but not the form of the metric. As an example of such a gauge
choice for a curved domain wall, consider Model II in
\cite{Behrndt:2002ee}. As the flow is along constant $\sigma $ and
$\theta =c\tau $ for constant $c$, the expression $q^{X\prime}
\omega_X^r$ has components $q^{X\prime} \omega_X^2=cq^{X\prime}
\omega_X^1$ and $q^{X\prime} \omega_X^3=0$. Hence it points only in one
direction, and though it is a complicated expression, an $SU(2)$ gauge
transformation in that one direction can annihilate $q^{X\prime}
\omega_X^r$. In the equations below, we will not explicitly use this
$SU(2)$ gauge choice. However, to reproduce the formulae of fake
supergravity such a gauge choice has to be assumed.

It remains to check the hyperino equation (\ref{realhyperino}), which we
already transformed to (\ref{BPS1}). Just as for the vector scalars in
the previous section, we can now again choose adapted coordinates on
$\mathcal{M}_{\rm Q}$ such that only one of the scalars $q^{X}$ has a
non-trivial $r$-dependence. We choose to call this scalar field $q$, and
denote the orthogonal, constant, scalars by $q^{\hat{X}}$:
\begin{equation}
g_{XY}= \left( \begin{array}{cc}
g_{qq} & 0 \\
0 & g_{\hat{X}\hat{Y}}
\end{array} \right), \label{adaptedmetric2}
\end{equation}
The supersymmetry condition (\ref{BPS1}) now splits into two equations:
\begin{eqnarray}
q^{\prime}\mathbf{\Theta} -\ifiTheta\rmi\,\else\fi g\,\sqrt{6}D_{q}\mathbf{P}&=&0,\label{hyper1}\\
q^{\prime}[\mathbf{R}_{\hat{X}q},\mathbf{\Theta}]-\ifiTheta\rmi\,\else\fi
g\,\sqrt{6} D_{\hat{X}}\mathbf{P}&=&0.\label{hyper2}
\end{eqnarray}
In view of (\ref{projepsilon}), the first equation (\ref{hyper1}) is
easily seen to be equivalent to the fake supergravity equation
(\ref{Freedman3}), \emph{provided} the $SU(2)$ gauge (\ref{SU2gauge}) has
been imposed.

The second equation (\ref{hyper2}), on the other hand, plays a somewhat
different r\^{o}le. First note that it is different from the
corresponding equation (\ref{hatP}) in Section \ref{ss:runvectors},
where we had only non-trivial vector scalars. In that case, the
derivative of $\mathbf{P}$ with respect to the orthogonal, constant
scalars $\varphi^{\hat{x}}$ had to vanish, whereas in the case of running
hyper-scalars, (\ref{hyper2}) no longer implies the independence of
$\mathbf{P}$
 of the orthogonal scalars
$q^{\hat{X}}$. In fact, squaring (\ref{hyper2}) and  using (\ref{hyper1})
and the quaternionic identity (\ref{quaterid}), one obtains,
 on a supersymmetric  flow solution,
\begin{equation}
D_{\hat{X}}\mathbf{P} D^{\hat{X}}\mathbf{P}=  2 D_{q} \mathbf{P} D^{q}
\mathbf{P}. \label{Phatq}
\end{equation}
This equation  shows that at least some of the $D_{\hat{X}}\mathbf{P}$
have to be non-zero and illustrates the meaning of (\ref{hyper2}), which
can be thought of as a constraint on the hatted derivatives of
$\mathbf{P}$ that allows one to effectively eliminate the dependence of
the equations on the constant scalars $q^{\hat{X}}$. The fact that this
``elimination'' of the $q^{\hat{X}}$ proceeds in a much less trivial way
than for the vector scalars $\varphi^{\hat{x}}$, has also  important
consequences for the scalar potential. Recalling that
$\partial_{x}\mathbf{P}=0$, the potential  (\ref{VPP2}) is
\begin{equation}
\mathcal{V}\unity =4 \mathbf{P}^2
-(D_X\mathbf{P})(D^{X}\mathbf{P})\label{VPP4}.
\end{equation}
Naively, this seems to have the wrong prefactor $(-1)$ instead of $(-3)$
in front of $(D_X\mathbf{P})(D^{X}\mathbf{P})$ in order to be
identifiable  with the  scalar potential (\ref{Freedman0}) of fake
supergravity. However, we can at most expect to identify these two
expressions after we expressed everything in terms of the only running
scalar $q$, and, indeed, (\ref{Phatq}) precisely corrects    the
prefactor $(-1)$ to $(-3)$:
\begin{equation}
\mathcal{V}\unity =4 \mathbf{P}^2
-3(D_q\mathbf{P})(D^{q}\mathbf{P})\label{VPP5}.
\end{equation}

Thus, in adapted coordinates, the supersymmetry conditions and the scalar
potential agree with the corresponding expressions in fake supergravity,
\emph{provided} the gauge choice (\ref{SU2gauge}) is taken. As the
$SU(2)$ curvature is proportional to the hypercomplex structure, and
hence non-degenerate, $\partial_{q}\omega_{\hat{X}}^{r}$ has to be
non-zero on the flow curve in the gauge where $\omega_{q}^{r}=0$. These
non-vanishing components are important for the consistency of
(\ref{hyper2}) with
 (\ref{Phatq}).

As for the consistency condition (\ref{commutatorD}), which for hypers
only reads
\begin{equation}
  \left[ q^{X\prime}D_{X}\mathbf{ P}\,,\, %q ^{X\prime\prime}D_X\mathbf{P}+
    q ^{X\prime}D_X q
  ^{Y\prime} D_Y \mathbf{P}  + \frac{1}{3} q^{Y\prime}q^{Z\prime} g_{YZ}
  \mathbf{P}\right]  =0,
 \label{commutator3}
\end{equation}
the use of adapted coordinates yields
\begin{equation}
\left[ D_{q } \mathbf{P}  , D_{q} D_{q} \mathbf{P}
+\frac{1}{3}g_{qq}\mathbf{P}
 \right] =0. \label{commutator10}
\end{equation}
Again, this is equivalent to the fake supergravity equation
(\ref{Freedman16}) \emph{provided} the $SU(2)$ gauge (\ref{SU2gauge}) is
adopted. Note also that in contrast to the vector scalars, the
hyper-scalars do not, in general, have to satisfy an analogue of the very
special geometric identity (\ref{veryspecialidentity}) that could render
the compatibility condition (\ref{commutator10}) automatically
inconsistent for curved domain walls. In fact, it is known that curved
BPS-domain walls supported by hyper-scalars exist
\cite{Cardoso:2002ec,Behrndt:2002ee}.

To sum up, we have shown that a curved BPS domain wall supported by a
hyper-scalar falls into the framework of  fake supergravity,
\emph{provided that} the $SU(2)$ gauge (\ref{SU2gauge}) is imposed.

%%%%%%%%%%%%%%%%%%%%%%%%%%%%%%%%%%%%%%%%%%%%%%%%%%%%%%%%%%%%%%%%%%%%%

%Section 4.3: Running vector and hypermultiplets

%%%%%%%%%%%%%%%%%%%%%%%%%%%%%%%%%%%%%%%%%%%%%%%%%%%%%%%%%%%%%%%%%%%%

\subsection{Running vector- and hyper-scalars} \label{ss:vectorandhyper}

We conclude the comparison between the supergravity and the fake
formalism by studying the case of non-trivial vector- and hyper-scalars.
Applying the experience gained in the previous sections, we will show
that also this general case can, at least locally, be included in the
formalism of fake supergravity.  This requires the choice of the adapted
coordinates in two steps.  First, we move to a coordinate system in which
just one scalar of the vector multiplet and one hyper-scalar are running,
namely $\varphi$ and $q$. According to the results of the previous
section, in this step it is necessary to adopt the $SU(2)$ gauge
(\ref{SU2gauge}) that removes the $SU(2)$ spin-connection from the
expression $D_r$, and to cast the hyperini equation (as well as its
corresponding contribution to the potential) in the same form as the
gaugini equation (and its corresponding contribution to the potential).
In this way, the two sectors become in many aspects analogous, as will be
explained more thoroughly in the next section. Here, we only focus on the
commutator constraint (\ref{commutatorD}), which, in these adapted
coordinates on $\mathcal{M}_{\rm VS}$ and $\mathcal{M}_{\rm Q}$ reduces
to
\begin{equation}
  \left[\partial_r \mathbf{ P} \,,\, \partial_r\partial_r\mathbf{P}
  + \frac{1}{3} \left(g_{qq}(q')^2 + g_{\varphi\varphi} (\varphi')^2\right)\mathbf{P}\right]  =0,
 \label{commutatorvh}
\end{equation}
where $\partial_r=q^\prime\partial_q + \varphi^\prime \partial_\varphi$.
Note that  there is no mixing of kinetic terms of vector- and
hypermultiplets, hence $g_{\varphi q}=0$.

We can now perform a second change of coordinates in order to have just
one scalar flowing, which is a combination of the scalars of the two
sectors. Normally, coordinate transformations that mix vector and
hypermultiplet scalars completely obscure the supersymmetry of a
supergravity theory. In our reduced and gauge fixed system of equations,
however, both types of scalars enter symmetrically, and one can consider
non-trivial coordinate transformations in the plane $(\varphi,q)$. We can
then take a new,
``total'', adapted coordinate system, in which only one scalar field
$\phi(r)$ is running, whereas the other one, which we will call
$\hat{\phi}$, is constant and orthogonal to $\phi$, at least along the
flow. Thus, we use the coordinate transformation
\begin{equation}
(\varphi(r),q(r)) \rightarrow (\phi(r),\hat{\phi}),
 \label{coordtrphi}
\end{equation}
with $\partial_{r}=\phi^{\prime}\partial_{\phi}$. Dropping the vanishing
terms and the overall factors in the commutator as in section
\ref{ss:runvectors}, we end up with
\begin{equation}
  \left[\partial_\phi \mathbf{ P} \,,\, \partial_\phi\partial_\phi\mathbf{P}
  + \frac{1}{3} g_{\phi\phi} \mathbf{P}\right]  =0.
 \label{commutatorvh2}
\end{equation}
Now, setting $g_{\phi\phi} =1$ by rescaling $\phi $, the above commutator
reduces to the corresponding expression (\ref{Freedman16}) of fake
supergravity.

We have here identified the commutator relation of fake supergravity,
which is a consistency condition of the BPS equations and the potential.
Our task of the next section will be to identify these BPS equations and
to show how the potential of true supergravity reduces to the one of fake
supergravity such that the identification of this commutator relation can
be understood.

We like to complete our discussion emphasizing that there is no
obstruction to the existence of curved domain walls in the presence of
non-trivial hypermultiplets and vector multiplets. In section
\ref{ss:runvectors} we showed that there can be no \emph{curved} BPS
domain walls that are supported solely by vector scalars. On the other
hand, there are known examples of AdS-sliced domain walls that are
supported by both vector and hyper-scalars \cite{Cardoso:2002ff}. One
might therefore wonder what exactly it is that the hypermultiplets do in
order to circumvent the
``no-go theorem'' for the vector multiplets. The material we have
accumulated in the previous sections allows us to give a simple answer to
this question.

In~(\ref{commutator1}) we have now
\begin{equation}
D_r\mathbf{\Theta}\equiv [\varphi^{x\prime} \, \partial_{x} +q^{X\prime}
 D_{X}]\mathbf{\Theta}.
\end{equation}
Inserting (\ref{Thetavector}) for $\mathbf{\Theta}$ into
(\ref{commutator1}) and dropping all terms that do not contribute to the
commutator, one derives
\begin{eqnarray}
0&=& \left[
\varphi^{x\prime}\partial_{x}\mathbf{P},D_{r}(\varphi^{x\prime}\partial_{x}\mathbf{P})+
\frac{1}{3}\varphi^{y\prime}\varphi^{z\prime}g_{yz}
\mathbf{P} \right]   \nonumber\\
&=&  \left[
   \varphi^{x\prime}\partial_{x}\mathbf{P}, \varphi^{x\prime\prime}\partial_{x}\mathbf{P}
   +\varphi^{x\prime}\varphi^{y\prime}\partial_{y}\partial_{x}\mathbf{P}+
\varphi^{x\prime}q^{X\prime}\partial_{x}D_{X}\mathbf{P}+\frac{1}{3}
\varphi^{y\prime}\varphi^{z\prime}g_{yz}\mathbf{P}     \right]
\end{eqnarray}
where $[\partial_{x},D_{X}]=0$ has been used.

Choosing again adapted coordinates $\varphi^{x}$ and $q^{X}$ such that
only one component of the $\varphi^{x}$ (which we call $\varphi$) and one
component of $q^{X}$ (which we call $q$) depends on $r$, the above
commutator simplifies to
\begin{equation}\label{newcommutator}
 \left[ \partial_{\varphi} \mathbf{P},  \partial_{\varphi}
\partial_{\varphi}\mathbf{P} +\frac{q^{\prime}}{\varphi^{\prime}}
\partial_{\varphi}D_{q}\mathbf{P}+\frac{1}{3}g_{\varphi\varphi} \mathbf{P}
 \right] = 0.
\end{equation}
Equation (\ref{consistency}) also simplifies to
\begin{equation}\label{consistency2}
\frac{D_{q}\mathbf{P}}{q^{\prime}g_{qq}}=\frac{\partial_{\varphi}\mathbf{P}}{\varphi^{\prime}g_{\varphi\varphi}}.
\end{equation}
One might now be tempted to use (\ref{consistency2})
 to re-express  $D_{q}\mathbf{P}$ in terms of
$\partial_{\varphi}\mathbf{P}$ in the commutator equation
(\ref{newcommutator}). Just as in section \ref{ss:runvectors}, one would
then again conclude that the only way to satisfy that constraint would be
by $[\partial_{\varphi} \mathbf{P},\mathbf{P}]=0$, which would then
forbid
curved domain walls.

However, there is a flaw in this argument: (\ref{consistency2}) is valid
only on the chosen flow curve, as it is based on a coordinate choice that
is adapted to that particular curve. Differentiating this equation with
respect to $\varphi$, however, probes this relation in a direction which
is not tangential to the curve, because we also have running
hyper-scalars. Away from the curve, however, (\ref{consistency2}) is in
general no longer valid. Thus, it is illegitimate to transform the mixed
derivative in (\ref{newcommutator}) into a pure $\varphi$-derivative
using a $\varphi$-derivative of  (\ref{consistency2}). What circumvents
the
``no-go theorem'' for vector scalars, is thus the presence of the mixed
derivative in (\ref{newcommutator}) and therefore this is how
hyper-scalars cure the incompatibility between curved walls and running
vector scalars.

%%%%%%%%%%%%%%%%%%%%%%%%%%%%%%%%%%%%%%%%%%%%%%%%%%%%%%%%%%%%%

\section{Similarities between vector and hypermultiplets}
\label{ss:similarities}

%%%%%%%%%%%%%%%%%%%%%%%%%%%%%%%%%%%%%%%%%%%%%%%%%%%%%%%%%%%%%

For a generic field configuration of $5D$ supergravity, the scalars of
vector and hypermultiplets  enter the field equations and the
supersymmetry transformation rules in a rather different way, due to the
distinct geometric structures of the corresponding scalar manifolds. This
is also  true for curved BPS-domain wall solutions when a generic
coordinate system $(\varphi^{x},q^{X})$ of the scalar manifold is used.
Indeed, the original papers on curved domain walls in $5D$ supergravity
\cite{LopesCardoso:2001rt,Cardoso:2002ec,Behrndt:2002ee,Cardoso:2002ff,Chamseddine:2001hx}
find visibly   different BPS equations for vector and hypermultiplet
scalars, and also the scalar potential does not appear
``symmetric'' with respect to vector and hyper scalars, as happens
for flat domain walls. In  sections
\ref{ss:runvectors}--\ref{ss:runhyperscalars}, on the other hand, we have
seen that the use of \emph{adapted} coordinates
$\varphi^{x}=(\varphi,\varphi^{\hat{x}})$ and $q^{X}=(q,q^{\hat{X}})$ and
the gauge fixing of the $SU(2)$ connection formally lead to  the same
expressions for both types of scalars in a BPS-domain wall background. As
this is an interesting result in its own right, we devote this extra
section to this observation and show explicitly how the adapted
coordinates lead to the same equations for both types of scalars also in
the formulation of
\cite{Cardoso:2002ec,Cardoso:2002ff,Chamseddine:2001hx}, where the BPS
equations are not expressed in the $SU(2)$ matrix-valued form of
(\ref{vectorBPS1})--(\ref{usefulform}).

The expressions in true supergravity that we are interested in,  are the
scalar potential
\begin{equation}
\mathcal{V}\unity =4 \mathbf{P}^2 -3
(\partial_{x}\mathbf{P})(\partial^{x}\mathbf{P})
-(D_X\mathbf{P})(D^{X}\mathbf{P}),\label{VPP7}
\end{equation}
and the matter BPS-equations (\ref{vectorBPS1}), (\ref{BPS1}),
\begin{eqnarray}
g_{yx}\varphi^{x\prime}\mathbf{ \Theta} - \ifiTheta\rmi\else\fi
\sqrt{6}g\partial_{y}\mathbf{P}&=&0, \label{vectorBPS3}\\
g_{YX}q^{X\prime}\mathbf{\Theta}+q^{X\prime}[\mathbf{R}_{YX},\mathbf{\Theta}]
-\ifiTheta\rmi\else\fi\sqrt{6}g  D_{Y}\mathbf{P} &=&0.  \label{BPS9}
\end{eqnarray}
Obviously, these expressions treat the vector- and the hyper-scalars
differently. On the other hand, from the results of the previous section,
we should be able to express them in a more symmetric form.

Let us first see, how the similarity between vector- and hyper-scalars
arises at the level of  the BPS-equations. As seen   in Sections
\ref{ss:runvectors} and \ref{ss:runhyperscalars}, using  adapted
coordinates, the BPS-equations (\ref{vectorBPS3}) and (\ref{BPS9})
simplify to
\begin{eqnarray}
\varphi^{\prime}\mathbf{\Theta}&=&\ifiTheta\rmi\else\fi\sqrt{6}\, g
\,g^{\varphi \varphi }
\partial_{\varphi} \mathbf{P},\label{vector1c}\\
0&=&\ifiTheta\rmi\else\fi\sqrt{6}\,g\, \partial_{\hat{x}}\mathbf{P},\label{vector2c}\\
 q^{\prime}\mathbf{\Theta}&=& \ifiTheta\rmi\else\fi\sqrt{6}\, g\,g^{qq} D_{q}\mathbf{P}, \label{hyper1c}\\
q^{\prime}[\mathbf{R}_{\hat{Y}q},\mathbf{\Theta}]&=&\ifiTheta\rmi\else\fi\sqrt{6}\,g\,
D_{\hat{Y}}\mathbf{P}.\label{hyper2c}
\end{eqnarray}
Modulo the $SU(2)$ connection, which can be gauged away along the flow
curve, (\ref{vector1c}) and (\ref{hyper1c}) have the same form. Moreover,
after the transformation (\ref{coordtrphi}) only one scalar is flowing
and, using the gauge (\ref{SU2gauge}), we have the new BPS-equations
\begin{eqnarray}
\phi^{\prime}\mathbf{\Theta}&=&\ifiTheta\rmi\else\fi\sqrt{6} g
g^{\phi\phi}
\partial_{\phi}\mathbf{P}\nonumber\\
0&=&\ifiTheta\rmi\else\fi\sqrt{6}g
g^{\hat{\phi}\hat{\phi}}\partial_{\hat{\phi}}\mathbf{P}. \label{vechyp5}
\end{eqnarray}
In this form, the BPS equation of the flowing scalar is the same as in
the fake supergravity theory.

The scalar potential, \emph{on a BPS-domain wall}, can also be made
symmetric between vector and hyper-scalars.
%and assume the simple form of (\ref{eq:Freedman44}).  XXX That equation
% comes later in the new setup.
The restriction to BPS-domain walls is crucial here, because proving
these statements requires using the information encoded in  the
orthogonal BPS equations (\ref{vector2c}) and (\ref{hyper2c}). Indeed, as
we saw in sections \ref{ss:runvectors} and \ref{ss:runhyperscalars},
eqs.~(\ref{vector2c}) and (\ref{hyper2c}) are constraints
 that allow one to eliminate the hatted derivatives
of $\mathbf{P}$ in the scalar potential (\ref{VPP7}) to obtain the
symmetric form
\begin{equation}
\mathcal{V}\unity =4 \mathbf{P}^2 -3 g^{\varphi \varphi
}(\partial_{\varphi }\mathbf{P})^2
-3g^{qq}(D_q\mathbf{P})^{2}.\label{VPP8}
\end{equation}
The gauge fixing of the $SU(2)$ connection and the transformation
(\ref{coordtrphi}) further simplify this to [using
$\partial_{\hat{\phi}}\mathbf{P}=0$ from (\ref{vechyp5})]
\begin{equation}
\mathcal{V}\unity = 4 \mathbf{P}^2 -3
g^{\phi\phi}(\partial_{\phi}\mathbf{P})^2,
\end{equation}
which reproduces (\ref{Freedman0}) of fake supergravity upon the
normalization $g^{\phi\phi}=1$.

%%%%%%%%%%%%%%%%%%

\medskip

We have now shown that the BPS equations and scalar potential can be put
in a form which treats symmetrically vector- and hyper-scalars when using
$SU(2)$ matrix-valued expressions. %\footnote{One should strictly speaking
%check whether (\ref{VPP8}) still reproduces the equations of motion, as
%it is the potential on the flow line in the original field space. As this
%line is itself the result of equations of motion%(or rather the BPS-equations)
%, however, what we have basically done is integrating out
%the other fields according to the usual effective field theory picture.}
In what follows we want to show that one can obtain more from the choice
of adapted coordinates and put also the BPS equations and potential
provided in \cite{Cardoso:2002ec,Cardoso:2002ff,Chamseddine:2001hx} in a
symmetric form. When this happens, we expect the BPS equations and
potential to match those of fake supergravity in \cite{DeWolfe:1999cp}.

Using the norm of $\bf W$, defined as\footnote{The unnatural factor in
this equation is due to the merging of the matrix notation in
\cite{Freedman:2003ax} and the notations in previous papers on true $5D$
gauged supergravity, where $g$ denotes the gauge coupling.}
$\mathbf{W}^2=\ft14\, g^2\, W^2 \,\unity$, the potential and first order
equations of fake supergravity become those of \cite{DeWolfe:1999cp}.
More precisely, the potential reads
\begin{equation}
V=g^2\,{\cal V},\qquad {\cal V}= - 6 \, W^2+\frac{9}{2} \, \gamma^{-2}\,
\partial_\phi
 W\partial_\phi W,
\label{eq:Freedman44}
\end{equation}
with $\gamma$ as in (\ref{Freedman12}), and the warp factor and scalar
field satisfy the first order equations
\begin{eqnarray}
U^{\prime} & = & \pm g \, \gamma\,  W ,\nonumber\\
\phi^{\prime} & = & \mp 3 g \, \gamma^{-1}\, \partial_{\phi} W.
\label{eq:phiprime}
\end{eqnarray}

Trying to mimic this form in real supergravity and following the ideas in
\cite{Ceresole:2001wi}, the authors of
\cite{Cardoso:2002ec,Cardoso:2002ff,Chamseddine:2001hx} split the
prepotential $\mathbf P$ in norm $W(\varphi , q)$ and phase $\mathbf Q
(\varphi , q)$
\begin{equation}
  P^r=\sqrt{\ft32}W Q^r,\qquad Q^rQ^r=1,\qquad \mbox{i.e.
  }\mathbf{Q}^2=-\unity.
 \label{splitPWQ}
\end{equation}
By doing so \cite{Ceresole:2001wi,LopesCardoso:2001rt,Cardoso:2002ec},
the potential gets closer to the fake supergravity one of
(\ref{eq:Freedman44}):
\begin{equation}
 {\cal V}= - 6 \, W^2+\frac{9}{2} \, \Gamma^{-2}\, g^{x y}\partial_x
 W\partial_y W\,
+\frac{9}{2} \, g^{XY}\partial_X  W\partial_Y W.
 \label{pot2}
\end{equation}
Here\footnote{Defining the analogous object $\Gamma _H^{-2}$ with
derivatives to the scalars $q^X$ rather than to $\varphi ^x$ would lead
to $\Gamma_H ^{-2}=3$ by using (\ref{KillingP}), which in the present
notations implies $W D_XQ^r=2\varepsilon ^{rst}Q^s{\cal R}^t_{XY}\partial
^YW$. This gives an understanding of the difference between (\ref{pot2})
and (\ref{VPP2}).}
\begin{equation}
\Gamma^{-2} (\varphi,q) \equiv  1 + W^2 \, \frac{g^{xy} (\partial _x
Q^s)(\partial_y Q^s)}{g^{x y}\partial_x W\partial_y W } .
 \label{Gamma}
\end{equation}

Also the BPS equations can be extracted from the $SU(2)$-valued form by
applying the above decomposition of $\mathbf P$ and by using the
projector \cite{LopesCardoso:2001rt}
\begin{equation}
\rmi \gamma_5 \epsilon = \left[A(r) \; {\mathbf Q}  + B(r) \; {\mathbf
M}\right] \epsilon, \label{procmod}
\end{equation}
where, $\mathbf M$ is a field-dependent phase orthogonal to $\mathbf Q$
(i.e.~$\mathbf{M}^2=-\unity$ and $\{\mathbf{Q},\mathbf{M}\}=0$) and $A$
and $B$ are functions of $r$, which satisfy the consistency requirement
$A^2(r) + B^2(r) = 1$. This is obviously related to the projector
(\ref{projepsilon}) introduced in section \ref{realreview} as
\begin{equation}
  \mathbf{\Theta} =
  \ifiTheta
      {\rm i}\left( A\, \mathbf{Q} + B\, \mathbf{M}\right),
  \else
       A\, \mathbf{Q} + B\, \mathbf{M},
  \fi
 \label{splitTheta}
\end{equation}
and the components $A$, $B$ and $\mathbf M$ can be read off from
(\ref{Thetavector}) and (\ref{Thetahyper}) by  simple projections. An
alternative way of fixing these functions is via the consistency
conditions that follow  from the integrability equations of the gravitino
variation \cite{LopesCardoso:2001rt,Cardoso:2002ec,Behrndt:2002ee}. For
instance, an interesting relation that specifies $A$    in terms of a
function   of the cosmological constant on the domain wall follows from
the integrability of $\delta \psi_{m\, i}$ :
\begin{equation}
A = \mp \gamma (r) \equiv \sqrt{1+\frac{\rme^{-2U}}{L_{d}^{2} g^2 W^2}}.
\label{gamma}
\end{equation}
A further expression for $A$ may be obtained by the projection of
(\ref{vectorBPS1}) on $\mathbf{Q}$, which results in
\begin{equation}
g_{xy}\varphi ^{x\prime}=3 gA^{-1} \partial _yW. \label{intermediate1}
\end{equation}
The consistency of the square of (\ref{intermediate1}) with the square of
(\ref{vectorBPS1}) then yields
\begin{equation}
A^{-2}  =\frac{2}{3}\frac{\partial _x P^r\partial ^x P^r}{\partial
_yW\partial ^y W}= 1 + W^2 \, \frac{g^{xy} (\partial _x Q^s)(\partial_y
Q^s)}{ g^{x y}\partial_x
 W\partial_y W }=\Gamma ^{-2},
\end{equation}
which further implies that (\ref{Gamma}) must also satisfy $\Gamma = \mp
\gamma$ (so far, $\Gamma $ was only defined up to a sign). At this point
the other integrability conditions following from the gravitino
transformations are identically satisfied and one can write the BPS
equations of the system in terms of the scalar function $W$
\cite{Cardoso:2002ec,Chamseddine:2001hx}:
\begin{eqnarray}
U^{\prime} & = & \pm g \, \gamma \, W, \label{eq:Uprimereal}\\ [2mm]
\phi^{\Lambda\prime} & = & \mp 3 g\, G^{\Lambda\Sigma} \partial_{\Sigma}
W, \label{eq:phiprimereal}
\end{eqnarray}
where $G^{\Lambda\Sigma}$ is defined by
\begin{eqnarray}
G^{xy} & = & \gamma^{-1}\, g^{xy}, \label{eq:G1}\\ [2mm]
 G^{XY} & = & \gamma \,g^{XY} + 2 \sqrt{1-\gamma^{2}}\,
 \varepsilon^{rst} M^r Q^s R^{tXY}, \label{eq:G2}\\ [2mm]
G^{xY} &=& G^{Xy} = 0. \label{eq:G3}
\end{eqnarray}
Notice that when the domain wall becomes flat, i.e. when $\gamma = 1$,
the projector reduces to $\mathbf \Theta = \ifiTheta\rmi\else\fi \mathbf
Q$, $G^{\Lambda\Sigma}= g^{\Lambda\Sigma}$ and we recover the BPS
equations of \cite{Ceresole:2001wi}.

Eq.~(\ref{pot2}) and (\ref{eq:G1})--(\ref{eq:G3}) show explicitly the
afore-mentioned asymmetry between vector and hypermultiplets which
appears in the formulation of
\cite{Cardoso:2002ec,Cardoso:2002ff,Chamseddine:2001hx} (This is encoded
for instance in the different expressions for $G^{xy}$ and $G^{XY}$).
However, we are now in a position to show that this apparent asymmetry
disappears when one uses the adapted coordinates.

Let us start from the hyperino BPS equation. Contracting (\ref{hyper1c})
with $\bf \Theta$ and using the decomposition in norm and phase of
$\mathbf P$, one can write
\begin{equation}
q^{\prime} =\ifiTheta-3\rmi\else3\fi g\,g^{qq}\Theta^r \left[ W
D_{q}Q^{r}+ (\partial_{q}W) Q^{r}\right]. \label{eq:eq:pro1}
\end{equation}
The last term can be simplified by using $\Theta^r Q^{r} = \mp
\ifiTheta\rmi\else\fi\gamma$ [see (\ref{splitTheta}) and (\ref{gamma})]
while $\Theta^{r}D_{q}Q^{r}$ can be determined from equation (28) of
\cite{Cardoso:2002ec}, which reads
\begin{equation}
q^{\prime}B M^r D_{q}Q^{r} = \mp \left(\frac{1-\gamma^{2}}{\gamma
W}\right)q^{\prime}\partial_{q}W\,. \label{eq:28}
\end{equation}
Adding $ A Q^{r}$ to the left-hand side and recalling that
$Q^{r}D_{q}Q^{r}= 0$, this actually becomes
\begin{equation}
\ifiTheta-\rmi\else\fi  \Theta ^rD_{q}Q^{r} = \mp
\left(\frac{1-\gamma^{2}}{\gamma W}\right)\partial_{q}W.
 \label{valuethetatDq}
\end{equation}
Substituting these expressions for the projections $\Theta^r Q^{r}$ and
$\Theta ^rD_{q}Q^{r}$ into (\ref{eq:eq:pro1}) we finally obtain
\begin{eqnarray}
g_{qq}q^{\prime} &=& \mp 3 \,g\,\left(\frac{1-\gamma^{2}}{\gamma}\right)
 \partial_{q}W \mp 3 \,g\, \gamma \partial_{q}W
\nonumber\\
&=& \mp 3g\frac{1}{\gamma}\partial_{q}W.\label{eq:pro}
\end{eqnarray}
Using the inverse metric we finally get that (\ref{eq:pro}) takes the
same form \footnote{Eq.~(26) in \cite{Cardoso:2002ec} already proved the
contracted version of (\ref{eq:pro}).} as (\ref{eq:G1}) and that both
look like (\ref{eq:phiprime}). This shows that, in adapted coordinates,
(\ref{eq:phiprimereal}) implies the same form (\ref{eq:phiprime}) for
both vector and hyper-scalars despite the apparent asymmetry encoded in
the matrix $G^{\Lambda \Sigma}$.

Also the potential (\ref{pot2}) gets now a symmetric form using the
adapted coordinates. In order to show this, one uses the fact that the
non-vanishing of $D_{\hat{X}}\mathbf{P}$, implied by (\ref{Phatq}), has
some important consequences for the derivatives of $W=\sqrt{\frac{2}{3}
P^r P^r}$. Indeed, $D_{\hat{X}}\mathbf{P}\neq 0$  in general implies that
$\partial_{\hat{X}}W\neq 0$, and the true supergravity potential becomes
\begin{eqnarray}
\mathcal{V}&=&-6 W^2+
\frac{9}{2}\Gamma^{-2}(\partial_{x}W)(\partial^{x}W) +
\frac{9}{2}(\partial_{X} W)(\partial^{X} W)\nonumber\\
 &=& -6W^2 + \frac{9}{2}\Gamma^{-2}(\partial_{\varphi}W)(\partial^\varphi W) +
\frac{9}{2}{\tilde{\Gamma}}^{-2}(\partial_{q}W)(\partial ^q W),
\label{VWW}
\end{eqnarray}
where
\begin{eqnarray}
\Gamma^{-2}&=&
1+W^2\,\frac{g^{xy}(\partial_{x}Q^{r})(\partial_{y}Q^{r})}{g^{xy}(\partial_{x}W)
(\partial_{y}W)}=
1+W^2\,\frac{(\partial_{\varphi}Q^{r})(\partial^{\varphi}Q^{r})}
{(\partial_{\varphi}W)(\partial ^\varphi W)},
\label{Gamma3}\\
{\tilde{\Gamma}}^{-2}&=&1+\frac{\partial_{\hat{X}}W\partial^{\hat{X}}W}{(\partial_q
W)(\partial ^qW)},\label{gamma2}
\end{eqnarray}
with the last term in (\ref{gamma2}) possibly  non-zero.

To increase the similarity between these formulae, one recalls from
(\ref{Phatq})  that
\begin{eqnarray}
 2\,\mathcal{N}_{iA}\mathcal{N}^{iA}&=&
\frac{9}{2}\,\partial_{X}W\partial^{X}W = \frac{9}{2}\,
\left(\partial_{q}W\right)\left(\partial^{q}W\right)
+\frac{9}{2}\, (\partial_{\hat{X}} W) (\partial^{\hat{X}} W)\nonumber\\
&=&(D_{X}P^r)  (D^{X}P^r)=3 (D_{q}P^r)( D^{q}P^r)\nonumber\\
&=& \frac{9}{2}\, \left[ (\partial_{q}W)(\partial ^qW) +W^2 (D_{q}Q^r)
(D^{q}Q^r) \right],
\end{eqnarray}
so that
\begin{equation}
\partial_{\hat X}W\partial^{\hat X}W =
W^{2} D_q Q^r D^q Q^r,
\end{equation}
 and thus
\begin{equation}
{\tilde{\Gamma}}^{-2}=1+ W^{2}\,\frac{D_{q}Q^r D^q Q^{r}}{(\partial_q
W)(\partial ^qW)}, \label{eq:new3}
\end{equation}
 which is then exactly as for the vector scalars in (\ref{Gamma3}). Hence, again, we see that
the similarity with the vector scalars only appears \emph{after} the
``deactivated'' hatted hyper-scalars have been properly taken care of.
%YYY Again, I took out the sentence about not fixing the SU(2) gauge, as
%we clearly do have to fix it in order to make (\ref{eq:new3}) look like
%the vector analogue. YYY We also stress that there was no need here to
%gauge fix the $SU(2)$ connection.

Using (\ref{eq:new3}) in (\ref{VWW}) one gets a perfectly symmetric form
between vector- and hyper-scalars, but the potential is not yet exactly
in the form of (\ref{eq:Freedman44})
\begin{equation}
\mathcal{V}=-6\,W^2+\frac{9}{2}\,\gamma^{-2}\,(\partial_{\phi}W)^2,
\label{Freedman44copy}
\end{equation}
which contains only \emph{one} scalar field $\phi$. In order to reproduce
(\ref{Freedman44copy}), one first recalls that $\Gamma^2=\gamma^2$. A
similar relation can also be derived for the hypermultiplet analogue
$\tilde{\Gamma}$, by projecting (\ref{hyper1c}) on $\mathbf{Q}$, which
gives
\begin{equation}
q^{\prime}=3gA^{-1}g^{qq} \partial_{q}W.  \label{Wqproof}
\end{equation}
The consistency of the square of (\ref{Wqproof}) and the square of
 (\ref{hyper1c})   then implies
\begin{equation}
A^{-2}=1+ W^{2}\,\frac{D_{q}Q^r D^q Q^{r}}{(\partial_q W)(\partial ^qW)}=
{\tilde{\Gamma}}^{-2}
\end{equation}
and hence ${\tilde{\Gamma}}^2=\gamma^2$ via (\ref{gamma}). Thus,
(\ref{VWW}) becomes
\begin{equation}
\mathcal{V}=-6\,W^2+\frac{9}{2}\,\gamma^{-2}\, [
(\partial_{\varphi}W)(\partial ^\varphi W) +
(\partial_{q}W)(\partial^qW)].
\end{equation}
If the domain wall is supported by vector scalars ($\phi=\varphi$) or by
hyperscalars ($\phi=q$), we have $\partial_{q}W=0$ or
$\partial_{\varphi}W =0$, respectively, and (\ref{Freedman44copy}) is
reproduced. In the mixed case, one has to go to the total adapted
coordinates $(\phi(r),\hat{\phi})$, and also obtains
(\ref{Freedman44copy}). Note that in this formulation with $\gamma$
instead of $\Gamma$ and $\tilde{\Gamma}$, the explicit dependence on the
$SU(2)$ connection has disappeared from the scalar potential and the BPS
equations. It only reenters upon the identification of $\gamma^2$ with
${\tilde{\Gamma}}^2$.

Finally, we can also read off the conditions for a BPS domain wall to be
curved. If the domain wall is supported by vector scalars only, a domain
wall would be curved if $\Gamma\neq 1 \Leftrightarrow
  \partial_{\varphi}Q^r\neq 0$.
As we saw, however, this is incompatible with the constraints imposed by
very special geometry.

A BPS domain wall of true supergravity that is supported by hyper-scalars
only, by contrast, is curved if any of the following equivalent
conditions is satisfied (they are equivalent on a BPS-domain wall
solution):
\begin{equation}
\partial_{\hat{X}} W \neq 0
 \Leftrightarrow  D_{q} Q^r \neq 0.
\end{equation}

As there are examples of such domain walls in true supergravity, these
conditions, as well as the commutator constraint
\begin{equation}
\left[ D_{q}\mathbf{P}, D_{q} D_{q} \mathbf{P}  + \frac{1}{3}g_{qq}
\mathbf{P} \right]=0,
\end{equation}
have solutions in quaternionic geometry.

%%%%%%%%%%%%%%%%%%%%%%%%%%%%%%%%%%%%%%%%%%%%%%%%%%%%%%%%%%%%%%%%%

%Section9: Conclusions

%%%%%%%%%%%%%%%%%%%%%%%%%%%%%%%%%%%%%%%%%%%%%%%%%%%%%%%%%%%%%%%%

\section{Conclusions}
\label{ss:conclusions}

Our original motivation to find the relation between fake supergravity
and genuine  supergravity partially evolved into an independent    and
insightful general study of curved BPS-domain walls in $5D$,
$\mathcal{N}=2$ gauged supergravity. Completing and clarifying previous
work, we have derived several results that deserve interest in their own
right. Most importantly, we  showed that curved BPS-domain walls in true
supergravity require non-trivial profiles of scalars  that sit in
hypermultiplets. It is interesting to notice that a similar outcome was
obtained in order to construct domain--wall solutions interpolating
between minima of the scalar potential as argued in \cite{Behrndt:2000km}
and then proved in \cite{Ceresole:2001wi,Alekseevsky:2001if,
Behrndt:2001km}. This result is of general validity and is  independent
of the relation to fake supergravity.

In order to make contact with fake supergravity, a true supergravity
theory has to be subjected to two types of gauge fixings. The first one
has to do with the above-mentioned  observation  that a supersymmetric
curved domain wall must involve non-trivial hyper-scalars. These in turn
introduce the $SU(2)$-connection, where $SU(2)$ is the factor of the
holonomy group of quaternionic-K{\"a}hler manifolds. This  $SU(2)$-connection
is absent in the fake supergravity framework of \cite{Freedman:2003ax}.
The equations of fake supergravity can thus only be reproduced if a
particular $SU(2)$ gauge, (\ref{SU2gauge}), is chosen. We showed that,
locally,  this is always possible.

The second type of gauge fixing is a partial fixing of the coordinates on
the scalar manifold of true supergravity. That is, one has
 to use ``adapted coordinates'',  in which only one scalar field is
flowing in order to make contact with the  one-scalar formalism of
\cite{Freedman:2003ax}. Clearly, the identification of this scalar and
hence the adapted coordinate system depend on the particular domain wall
one is considering, and is in general only a local coordinate choice on
the scalar manifold. It may even be a local choice for part of the flow
only. Indeed, a flow line in the complete scalar manifold may return to
the same point of the manifold but flowing in a different direction. This
implies that at this later stage of the flow, one has to use different
adapted coordinates, although one is describing the same region on the
scalar manifold.

The  adapted coordinates can be viewed as an analogue of the ``free
fall'' reference frame of a freely falling body, which, as in general
relativity,   is certainly somewhat unsatisfactory due to the breaking of
the general coordinate invariance. However, as a technical device, this
coordinate choice is essential to make contact with the one-field
formalism of \cite{Freedman:2003ax}.

The identification of true and fake supergravity applies only on the line
of flow of a chosen domain wall in the scalar manifold. The formulae for
the corresponding potentials can be made equal due to a relation
(\ref{Phatq}) between derivatives of the prepotential in directions along
and orthogonal to the line. This equation is a consequence of the BPS
equations of scalars not considered in the one-scalar formalism of fake
supergravity \cite{Freedman:2003ax}. This is nothing but an illustration
of the obvious further richness of ordinary supergravity, which contains
more equations than what can be encoded into  its 'fake' counterpart.
These extra equations should determine how the line of flow is embedded
into the larger scalar manifold of the full supergravity theory. The
equivalence indeed only holds along such a line determined by the true
supergravity equations. Furthermore, the full supergravity theory gives
the expression for the triplet superpotential $\mathbf{W}$. This object
is not specified in fake supergravity, and an arbitrary expression for
$\mathbf{W}$ cannot necessarily arise from a true supergravity.

Finally, as a further interesting observation whose relevance goes beyond
the comparison with fake supergravity, we have shown that the careful
choice of  ``adapted coordinates'' and  the fine tuning of the $SU(2)$
connection allows to describe BPS flows in a formalism that treats
symmetrically vector and hypermultiplet scalars, both with respect to
their  equations  of motion and the prefactors in the potential
(\ref{VPP8}). In this way, at least for the purposes of this paper, a lot
of information can be encoded in the dynamics of a single flowing
(possibly  ``hybrid'') scalar field.

\subsection*{Acknowledgments}

%\medskip
%
%\textbf{Acknowledgements:}
We thank Klaus Behrndt, Gabriel Lopes Cardoso, Mirjam Cveti\v{c},
Riccardo D'Auria, Sergio Ferrara, Dan Freedman, Renata Kallosh, Jan Louis
and Martin Schnabl for interesting discussions. Anna C. and G.D. are
grateful to K.U. Leuven for hospitality during the early stages of this
project.

This work is supported in part by the European Community's Human
Potential Programme under contract HPRN-CT-2000-00131 Quantum Spacetime.
The work of A.Celi and A.V.P. is supported in part by the Federal Office
for Scientific, Technical and Cultural Affairs through the
"Interuniversity Attraction Poles Programme -- Belgian Science Policy"
P5/27. The work of M.Z. is supported by an Emmy-Noether-Fellowship of the
 German Research Foundation (DFG) (ZA 279/1-1).

\newpage
%%%%%%%%%%%%%%%%%%%%%%%%%%%

\appendix

\section{The SU(2) gauge choice}
\label{app:SU2choice}

The geometric structure of quaternionic manifolds is determined by
complex structures. The 3 complex structures $(J^r)_X{}^Y$ form a span,
which means that one can rotate them locally in the manifold, i.e.
depending on local functions $l^r(q)$, without changing the geometry:
\begin{equation}
  \delta _l (J^r)_X{}^Y=\varepsilon ^{rst} l^s (J^t)_X{}^Y.
 \label{dellJ}
\end{equation}
Also other vector quantities, such as the moment maps, rotate in the same
way under these $su(2)$-reparametrizations. The gauge field is the
connection $\omega _X^r$:
\begin{equation}
  \delta _l\omega _X^r=-\ft12\partial _Xl^r+\varepsilon ^{rst} l^s\omega
  _X^t.
 \label{dellomega}
\end{equation}
As mentioned in section \ref{ss:5dN2SG}, the curvature of this gauge
field ${\cal R}^r_{XY}$ is proportional to the complex structure
multiplied by the quaternionic-K{\"a}hler metric. Killing spinors transform
in the doublet representation, i.e.
\begin{equation}
  \delta _l\epsilon _i= l^r(\sigma ^r)_i{}^j\epsilon _j.
 \label{dellepsilon}
\end{equation}

Notice that these are not local spacetime gauge transformations, but
transformations on the description of the quaternionic structures, local
in the quaternionic-K{\"a}hler manifold. This is the gauge freedom that we
are fixing with the choice (\ref{SU2gauge}). Note that these gauge
transformations leave the quaternionic metric $g_{XY}$ invariant and are
thus compatible with the adapted coordinate choice
(\ref{adaptedmetric2}). More details can be found in
\cite{Bergshoeff:2003yy,VanProeyen:2004xt}.

We now consider the finite transformations rather than the infinitesimal
ones mentioned above. These transform the doublet spinors as
$\epsilon_{i} \rightarrow V_{i}{}^{j} \epsilon_{j}$ and the $SU(2)$
connection transforms as
\begin{equation}
\upomega_{X} \rightarrow {\tilde{\upomega}}_{X}
=-\mathbf{V}(\partial_X-\upomega_X)\mathbf{V}^{-1}. \label{SU2trafo}
\end{equation}
Let $q^{X}(r)$ be a curve on $\mathcal{M}_{Q}$ with starting point
$q^{X}_{0}=q^{X}(0)$. The path-ordered exponential  (``Wilson line'')
\begin{equation}
\mathbf{U}(q^X(r),q^{X}_{0})\equiv \mathcal{P} \left\{ \exp \left[
\int_{0}^{r} \upomega_{X}(q^{X}(\bar{r})) q^{X\prime}(\bar{r})
\rmd\bar{r} \right] \right\}
\end{equation}
satisfies
\begin{eqnarray}
\frac{\rmd}{\rmd
r}\mathbf{U}(q^{X}(r),q^{X}_{0})&=&q^{X\prime}(r)\upomega_X
(q^{X}(r))\mathbf{U}(q^{X}(r),q^{X}_{0})\nonumber\\
\Leftrightarrow q^{X\prime}(\partial_{X}- \upomega_X)\mathbf{U}&=&0.
\label{annih}
\end{eqnarray}
$\mathbf{U}$   has been defined  only on the curve, but there should be
some analytic continuation of that function, at least in a neighborhood
of the curve. If we now choose the $SU(2)$ transformation
\begin{equation}
\mathbf{V}=\mathbf{U}^{-1},
\end{equation}
on this neighborhood, the tangential component of  the new, gauge
transformed, $SU(2)$ connection (\ref{SU2trafo}) becomes,  remembering
(\ref{annih}),
\begin{equation}
q^{X\prime}{\tilde{\upomega}}_{X}=-q^{X\prime}
\mathbf{U}^{-1}(\partial_{X}-\upomega_X)\mathbf{U} =0.
\end{equation}
In other words, the   component of the $SU(2)$ connection $\upomega_X$
tangential to the flow curve can always be gauged away for sufficiently
short curve segments.

Using these notations, one may also rephrase the procedure of adopting
the gauge choice (\ref{SU2gauge}) as replacing the identification
(\ref{WP}) with
\begin{equation}
\mathbf{W} = -\frac{\rmi g}{\sqrt{6}}\mathbf{U}^{-1} \mathbf{P}
\mathbf{U},
 \label{WPU}
\end{equation}
in a patch where the flows do not intersect and starting from some
$q_0^X$ on any line of flow.

%\bibliography{refd5conf}
%\bibliographystyle{toine}
\providecommand{\href}[2]{#2}\begingroup\raggedright\endgroup

\end{document}